\documentclass[hidelinks,aps,pra,twocolumn,superscriptaddress,10pt]{revtex4-2} 
\usepackage{amsmath}
\usepackage{amssymb}
\usepackage{graphicx}
\usepackage{color}
\usepackage{url}
\newcommand{\be}[1]{\begin{equation}\label{#1}}
\newcommand{\ee}{\end{equation}}
\newcommand{\bea}[1]{\begin{eqnarray}\label{#1}}
\newcommand{\eea}{\end{eqnarray}}
\newcommand{\no}{\nonumber \\}
\newcommand{\Fig}[1]{Fig.(\ref{#1})}

\newcommand{\Eq}[1]{Eq.(\ref{#1})}
\newcommand{\App}[1]{Appendix~\ref{#1}}
\newcommand{\Sec}[1]{Section~\ref{#1}}
\newcommand{\bsub}{\begin{subequations}}
\newcommand{\esub}{\end{subequations}}
\newcommand{\bwt}{\begin{widetext}}
\newcommand{\ewt}{\end{widetext}}

\usepackage{color}

\def\trm#1{\textrm{#1}}
\def\tit#1{\textit{#1}}

\def\ttt#1{\texttt{#1}}

\def\a0{{\alpha_0}}

\def\da0{{\dot{\alpha}_0}}

\def\myoverDefn#1#2{\hbox{\space \raise-2mm\hbox{$\textstyle{#1} \atop \scriptstyle{#2}$} }}

\def\a{{\alpha}}

\def\dag{\dagger}

\def\rp{r_{P}}
\def\rp2{r_{p}^{2}}
\def\Tr{\textrm{Tr}}

\newcommand{\ket}[1]{|#1\rangle}
\newcommand{\bra}[1]{\langle #1|}

\newcommand{\IP}[2]{\langle {#1} | {#2} \rangle}
\usepackage{fancyhdr}
\begin{document}
\fancyhead[R]{\ifnum\value{page}<2\relax\else\thepage\fi}

\title{Extending the Hong-Ou-Mandel Effect: the power of nonclassicality}
\author{Paul M. Alsing}\email{corresponding author: paul.alsing@us.af.mil}
\affiliation{Air Force Research Laboratory, Information Directorate, 525 Brooks Rd, Rome, NY, 13411, USA}
\author{Richard J. Birrittella}
\affiliation{Air Force Research Laboratory, Information Directorate, 525 Brooks Rd, Rome, NY, 13411, USA}
\author{\\Christopher C. Gerry}
\affiliation{The City University of New York, Bronx, New York, 10468-1589,USA}
\author{Jihane Mimih}
\affiliation{Department of Electrical and Computer Engineering, Naval Postgraduate School, 1 University Circle, Monterey, California 93943, USA}
\author{Peter L. Knight}
\affiliation{Blackett Laboratory, Imperial College, London SW72AZ, UK}

\date{\today}

\begin{abstract}
We show that the parity (evenness or oddness) of a nonclassical state of light has a dominant influence on the interference effects at a balanced beam splitter, irrespective of the state initially occupying the other input mode.  Specifically, the parity of the nonclassical state gives rise to destructive interference effects that result in deep valleys in the output joint number distribution of which the Hong-Ou-Mandel (HOM) effect is a limiting case. The counter-intuitive influence of even a single photon to control the output of a beam splitter illuminated by any field, be it a coherent or even a noisy thermal field, demonstrates the extraordinary power of non-classicality. The canonical example of total destructive interference of quantum amplitudes leading to the absence of coincidence counts from a 50:50 beam splitter (BS) is the celebrated HOM effect, characterized by the vanishing of the joint probability of detecting singe photons in each of the output beams. We show that this is a limiting case of more general input states upon which a 50:50 BS can create total, or near total, destructive interference of quantum amplitudes. For the case of odd photon number input Fock state of arbitrary value $n>0$ we show that the joint photon number probabilities vanish when detecting identical photon numbers in each output beams. We specifically examine the mixing of photon number states of $n$ = 1, 2, and 3 with a continuous variable state, such as a coherent state of arbitrary amplitude, and a thermal state. These vanishing joint probabilities form what we call a central nodal line (CNL): a contiguous set of zeros representing complete destructive interference of quantum amplitudes. We further show that with odd or even photon number Fock states $n$ with $n>1$, that there will be additional “off-diagonal” curves along which the joint photon number probabilities are either zero, or near zero, which we call pseudo-nodal curves (PNC) which constitute a near, but not complete, destructive interference pattern in the photon number space. We interpret all of these interference effects as an extension of the HOM effect. We explain the origin of these effects and explore the experimental prospects for observing them with currently available number-resolving detectors in the presence of a small amount of noise. 
\end{abstract}

\maketitle
\thispagestyle{fancy}

\section{Introduction}\label{sec:Intro}
The generation of two-mode entangled states of light can be accomplished by mixing nonclassical single-mode states of light at a beam splitter (BS) \cite{Kim:2002}. 
The process that gives rise to such two-mode states of light via beam splitting is known 
as multiphoton interference 
\cite{Ou:2007},  
%
%
and serves as a critical element in several applications including quantum optical interferometry \cite{Pan:2012}, and quantum state engineering where beam splitters and conditional measurements are utilized to perform post-selection techniques such as photon subtraction \cite{Dakna:1997,Carranza:2012,Magana-Loaiza:2019}, photon addition  \cite{Dakna:1998}, and photon catalysis \cite{Lvovsky:2002, Bartley:2012, Birrittella:2018}.   

In spite of its name, ``multiphoton interference" does not involve the interference of photons. Rather, as has been emphasized by Glauber \cite{Glauber:1995}, 
it is always the addition of the quantum amplitudes (themselves being complex numbers) associated with these states that give rise to interference effects. 
The amplitudes to be added are those associated with different paths (or processes) to obtain a given final output state. Thus, the term “multiphoton interference” must be understood to mean “interference with states containing numerous photons.” The canonical example of this kind of interference is what has come to be known as the Hong-Ou-Mandel (HOM) effect \cite{HOM:1987}, 
which is a two-photon interference effect wherein single photons 
in either of the output beams 
of a 50:50 beam splitter emerge together (probabilistically). 
Detectors placed at each of the output ports will yield no simultaneous coincident clicks.
That is, the input state $\ket{1,1}_{ab}$   results in the output state  $\tfrac{1}{\sqrt{2}}\left( \ket{2,0}_{ab} + \ket{2,0}_{ab}\right)$. The absence of the  $\ket{1,1}_{ab}$ in the output is due to the complete destructive interference between the quantum amplitudes of the  two processes (both photons transmitted, or both reflected) that potentially would lead to the state  being in the output.
The essence of this effect from an experimental point of view is that the joint probability $P_{ab}(1,1)$  for detecting one photon in each of output beams vanishes, i.e.   $P_{ab}(1,1)=0$.

In this paper we show that the same complete destructive interference demonstrated by HOM effect persists for more generalized input states such that the joint probability for measuring equal number of photons at the output ports of a 50:50 BS vanishes.
These situations arise from the mixing of one-photon (Fock state (FS)) and a continuous variable (CV) state at the BS. 
That is, the input states could be 
$\ket{\Psi}_{ab}=\ket{1}_a\ket{\psi}_b$ or $\rho_{ab} = \ket{1}_a\bra{1}\otimes\rho_b$ where $\ket{\psi}_b$ and $\rho_{b}$ are CV pure and mixed states, respectively. For these situations we find that for a 50:50 BS the output probabilities $P_{ab}(m',m')=0$ for all integer
$m'\in\mathbb{Z}_{\ge0}$.
This means we obtain a \tit{central nodal line} (CNL), or a line of zeros representing complete destructive interference, along the diagonal of the output joint photon number distribution. We can understand this CNL 
(i.e. that  $P_{ab}(m',m')=0$) to be higher-order form of the HOM effect.
We further show that the mixing of odd photon number states with CV states, of the composite form 
$\ket{n}_{a}\bra{n}\otimes\rho_b$ with odd $n\in\mathbb{Z}^{odd}_{\ge0}$,
results in a CNL,  $P_{ab}(m',m')=0$, for arbitrary states $\rho_b$. 
%


The Hong-Ou-Mandel effect is often characterized (e.g. observed experimentally) by the \tit{dip} in the rate of coincident single-photon detections as a function of the position of the beam splitter (equivalently, the difference in the arrival times of the two single photons at the BS). We take the position that the whole of the dip is not, in and of itself, the central feature of the HOM \tit{effect}. Rather, the HOM effect is the quantum amplitude interference effect that occurs when two photons enter a beam splitter simultaneously from opposite sides. The  dip in the curve is the result of what needs to be performed experimentally to prepare and verify the required input state $\ket{1,1}_{ab}$. In the HOM experiment, the two-photons involved originate from the same source and adjusting the position of the beam splitter is required to ensure that the photons arrive there simultaneously. The HOM \tit{effect} is the complete destructive interference between two photons simultaneously  arriving (zero time delay) at opposite sides of a 50:50 BS, as opposed to it being the whole HOM  dip (e.g. a scan across the difference in arrival times between the two photons). In short, we distinguish the HOM \tit{effect} as the center (theoretical minimum) of the experimental \tit{dip}; that is, the point at which the output amplitude for coincident counts vanishes due to complete destructive quantum interference.

In the case of mixing a single photon with coherent light at a beam splitter, the photons involved are from independent sources. Furthermore, the coherent state is a CV state obtained from a phase-stabilized laser continuously being shone upon the BS. The single (signal) photon  will be known to have arrived at the BS by the heralding of its (idler) twin from a spontaneous parametric down-converter (SPDC). Thus, the mixing of the single photon with a coherent light beam will result in multiphoton interference in bursts conditioned by the heralding of the (signal) photon. Therefore, there will be no manifestation of an HOM dip in experiments of the type envisioned here. The canonical ($\ket{1,1}_{ab}$ input state) HOM dip occurs because there is no single photon continuous variable state that could be prepared and continuously shone upon a BS. 
It is the quantum amplitude interference effects themselves that are extensions of the HOM effect.

In addition to the diagonal CNL of zeros, for any $n>1$, even or odd,
there exists  sets of non-contiguous non-diagonal zeros lying on what we term \tit{pseudo-nodal curves} (PNC), or near-nodal curves, of the output joint probabilities that lie symmetrically placed about the diagonal. In the case of $n=2$ there are two non-diagonal PNC and no CNL, while for $n=3$ there are two non-diagonal PNC  that lie symmetrically placed on either side of the diagonal CNL. The more photons we mix on the BS, the more PNC we obtain. In fact, the mixing of the $n$-photon Fock state (FS) with a CV state, or a thermal state (TS) at a 50:50 BS results in the generation of $n$ PNC or CNL
(i.e. $n$ non-diagonal PNC for $n$ even, and $(n-1)$ PNC for $n$ odd - with the addition of one diagonal CNL). 
The effect of the PNC are to furcate (i.e. multi-divide) the output joint probability distribution into $(n+1)$ peaks with $n$ ``valleys." While these valleys are not true nodal curves, they do appear as minima in the output joint probability distribution since actual collections of (non-contiguous) zeros lie along these PNC.
All of these generalized quantum interference effects, the CNL and the PNC, 
are what we refer to as the \tit{extended HOM effect}.
\begin{figure}[ht]
\begin{center}
\hspace*{-0.20in}
\includegraphics[width=4.0in,height=1.75in]{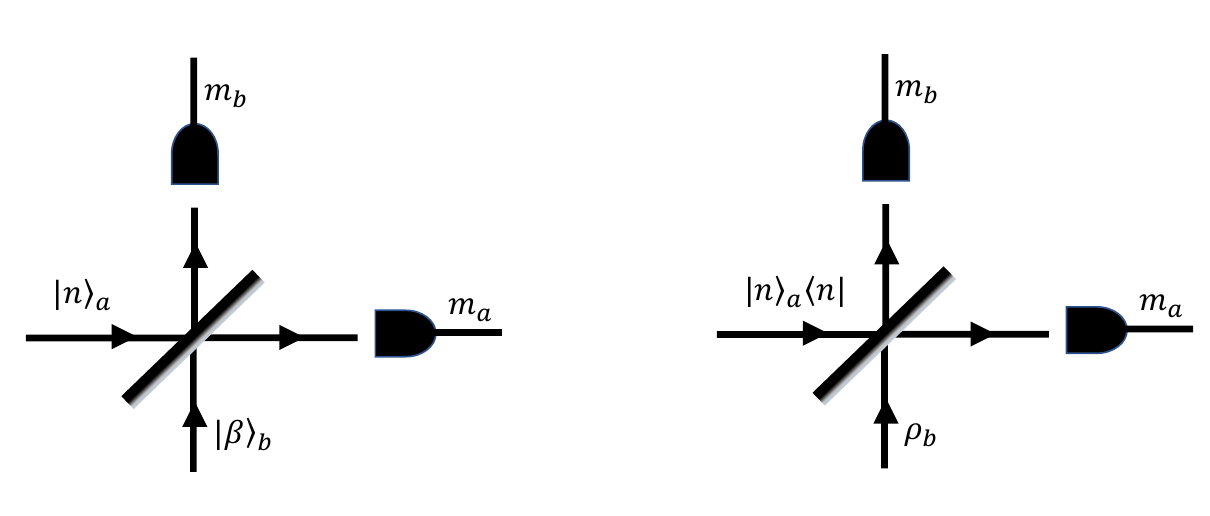}
\caption{\label{fig:FSCS:FSTS:cartoon} 
The measurement of $m_a$ and $m_b$ photons in the output ports of a BS of transmissivity $T=\cos^2(\theta/2)$ with 
input Fock state (FS) $\ket{n}_a$ and 
(left) coherent state (CS) $\ket{\beta}_b$, 
(right) thermal state (TS) $\rho^{thermal}_b$.
}
\end{center}
\end{figure}

An important point to mention here is that these effects (CNL and PNC) are \tit{independent} of the level of excitation of the CV states. In fact these effects  are fundamental intrinsic properties of the BS itself, 
particularly when used in a 50:50 configuration,
when acting upon discrete non-classical (Fock) states (and hence, states built up from these component states).
Specifically, 
the appearance of the CNL and PNC are independent of the amplitudes 
that describe the $a$- and $b$-mode input states, 
as long as the there are only odd numbered FS entering the $a$-mode, i.e. as long as the one of the states entering the BS is state of odd-parity \cite{BAG_Parity:2021,odd:parity:comment}, a definitively non-classical state.

The above effects involving a CV state, specifically a coherent state (CS) 
mixed with a $n$-photon FS on a 50:50 BS 
 \Fig{fig:FSCS:FSTS:cartoon} (left),
\begin{figure*}[th]
\begin{center}
\begin{tabular}{ccc}
\includegraphics[width=3.0in,height=1.75in]{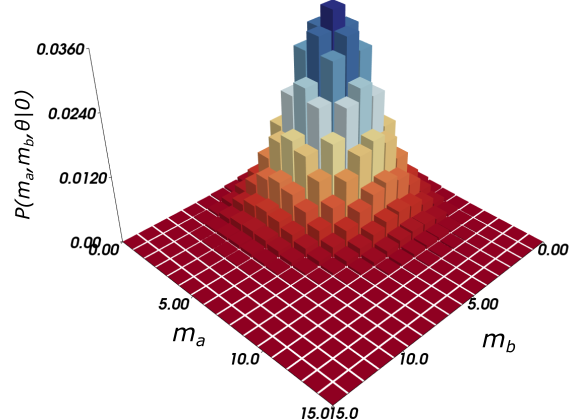}   & {} &
\includegraphics[width=3.0in,height=1.75in]{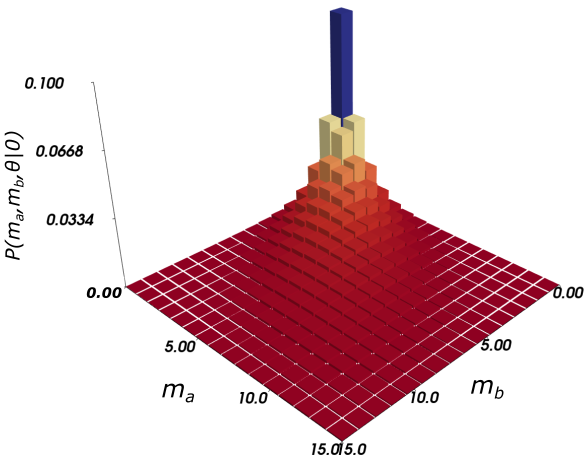}           \\
\includegraphics[width=3.0in,height=1.75in]{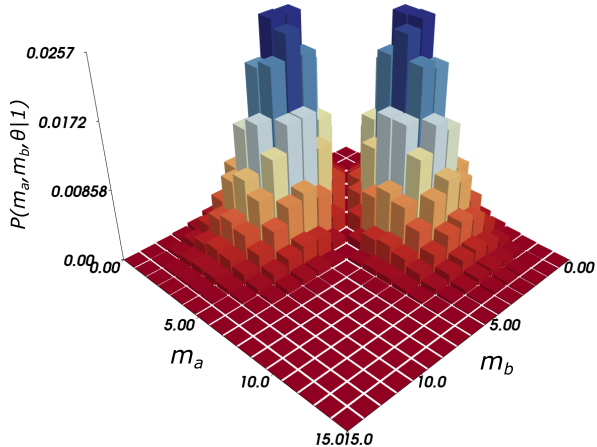}    &  {} &
\includegraphics[width=3.0in,height=1.75in]{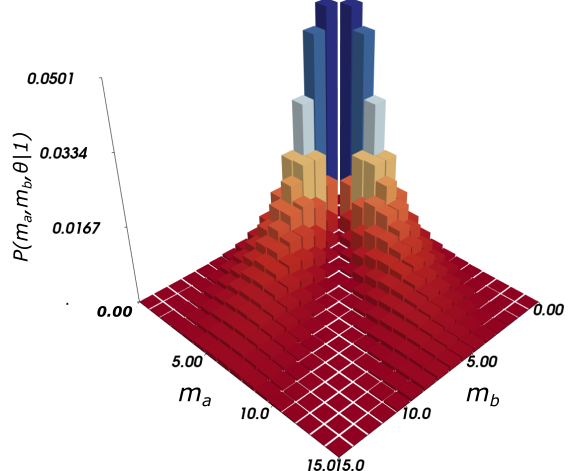}     
\end{tabular}
\caption{\label{fig:FSCS:FSTS:plots:n0:1} 
Diagonal central nodal line (CNL) and off-diagonal pseudo-nodal curves (PNC) 
for output probability $P(m_a, m_b|n)$ for 
(top row) $n=0$, (bottom row) $n=1$,
for measurement of  $m_a$ and $m_b$ photons 
at the output ports of a 50:50 BS for 
input Fock state (FS) $\ket{n}_a$ and 
(left column) coherent state (CS) $\ket{\beta}_b$ with average photon number $\bar{n}=9$, 
(right column) thermal state (TS) $\rho^{thermal}_b$ with average photon number $\bar{n}=9$,
as depicted in \Fig{fig:FSCS:FSTS:cartoon}.
}
\end{center}
\end{figure*}
\begin{figure*}[th]
\begin{center}
\begin{tabular}{ccc}
\includegraphics[width=3.0in,height=1.75in]{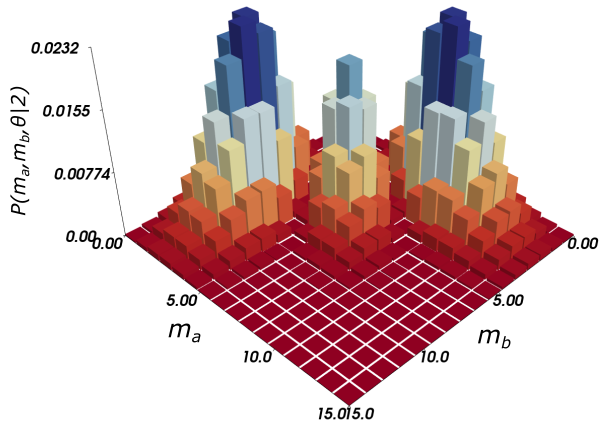}    &  {} &
\includegraphics[width=3.0in,height=1.75in]{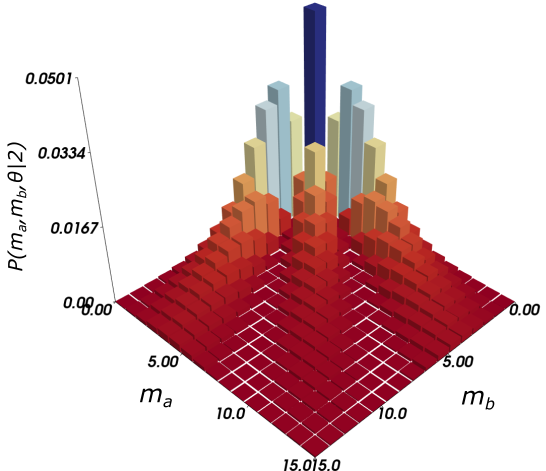}           \\
\includegraphics[width=3.0in,height=1.75in]{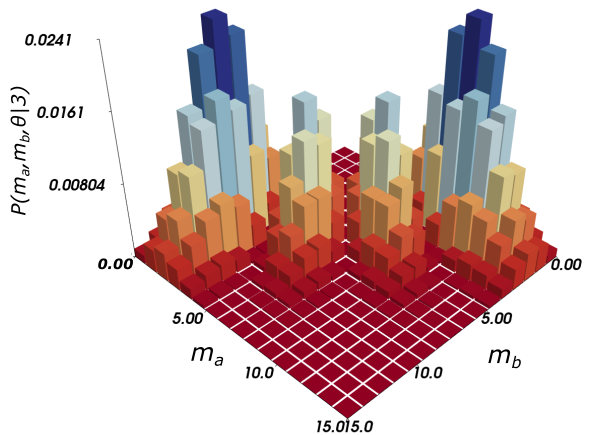}    & {} &
\includegraphics[width=3.0in,height=1.75in]{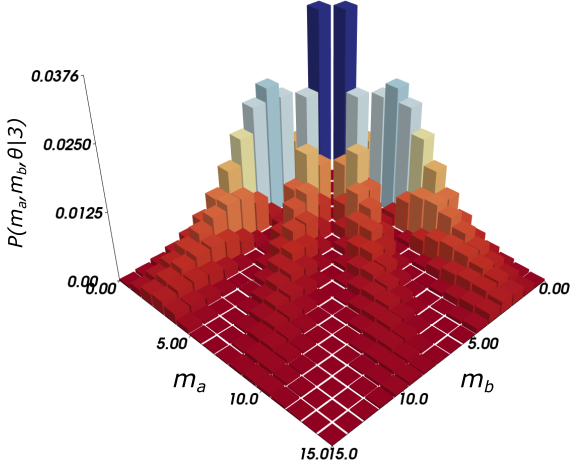} 
\end{tabular}
\caption{\label{fig:FSCS:FSTS:plots:n2:3} 
Same as \Fig{fig:FSCS:FSTS:plots:n0:1} 
(left column: CS; right column: TS)
now with (top row) $n=2$ and (bottom row) $n=3$.
}
\end{center}
\end{figure*}
were reported some years ago by Birrittella, Mimih and Gerry (BMG) \cite{BMG:2012}, and are shown in the 
left columns of \Fig{fig:FSCS:FSTS:plots:n0:1} and \Fig{fig:FSCS:FSTS:plots:n2:3}.
The FS/CS case investigated by BMG illustrates the above CNL and PNC features for the case of $n=1,2,3$, and serves as an archetypal example.
 The right columns of \Fig{fig:FSCS:FSTS:plots:n0:1} and \Fig{fig:FSCS:FSTS:plots:n2:3}
 illustrates the analogous CNL and PNC for the situation of a Fock state/Thermal state (FS/TS) input to a 50:50 BS as illustrated in the right cartoon of \Fig{fig:FSCS:FSTS:cartoon}.
 The similarity of the two cases illustrated in  \Fig{fig:FSCS:FSTS:plots:n0:1} and \Fig{fig:FSCS:FSTS:plots:n2:3}
  is readily apparent and exemplifies the universal behavior of the CNL and PNC as interference patterns in the space of photon numbers arising from the 50:50 BS.
The focus of BMG  was 
devoted to quantum interferometry, and the CNL and PNC effects were not fully explored in that work.
The goal of this present work is to fully explore these effects both analytically and numerically for general states and place them in the context of previous work on multiphoton interference effects.

As mentioned previously, the HOM effect \cite{HOM:1987} was the first discovered effect of this type in the laboratory although there were hints of such an effect in the theoretical work by other workers at around the same time. The history of the HOM effect and many of its ramifications has recently been reviewed by Bouchard \tit{et al.} \cite{Bouchard:2021}. In 1996, Ou \cite{Ou:1996} studied the multiphoton interference obtained by a single-photon mixing with an $n$-photon number states at 50:50 beam splitter, where it was shown that a strong destructive interference effect was manifested in the photon number distribution in the output. For an $n$-photon number state mixing with the vacuum the joint photon number distribution of the output state is a two-mode binomial. 
With the mixing of one photon with the $n$-photon state one finds, if $n$ is odd, that the joint probability for detecting  $(n+1)/2$ photons in each mode is zero, i.e.  $P_{ab}\big((n+1)/2, (n+1)/2\big)=0$,
as a consequence of quantum interference.
However, if $n$ is even, the combinatorics of mixing with one photon simply does not allow for output states with equal photon numbers in each mode to yield $P_{ab}(m,m)=0$ for all $m$ for those cases. In the same paper, Ou studied the mixing of a single photon with coherent light and with thermal light at a 50:50 beam splitter and noted that similar interference effects could be observed by balanced homodyne detection. 
Subsequently Kuzmich \tit{et al.} \cite{Kuzmich:1998} demonstrated this effect in the laboratory for coherent light mixing with a single photon. 

Ou \cite{Ou:1996} and Kuzmich \tit{et al.} \cite{Kuzmich:1998} studied only the case where one photon is mixed with a CV state at a beam splitter. Though they noted the important interference effects, they did not explicitly describe the existence of a nodal line of zeros representing complete destructive interference  in the joint photon number distribution for the output fields as was done by BMG [8].  Furthermore, the former authors did not extend their considerations to mixing CV states of light with $2$ or more photons, as was done by BMG \cite{BMG:2012}. 
Rarity, Tapster and Loudon \cite{Rarity:2005}  experimentally explored the HOM effect for the mixing of a single photon state with a separate weak coherent state and  showed (via the HOM dip procedure, see their Fig.(4)) the probability for observing one photon in each beamsplitter output approaches zero (at the minimum of the dip) due to destructive interference.
Recently, Podoshvedov and An \cite{Podoshvedov:2021} have proposed the generation of even/odd CV states by 
quantum interference of CV states with a delocalized photon (occupying two different spatial modes) on a beam splitter.
The emphasis of their work was on the entanglement properties of such generated states, as opposed to interference effects as discussed in this work.

The paper that we have found that comes closest to the spirit of this present work is that 
of
Lai, Bu\u{z}ek and Knight (LBK) \cite{Lai:1991} which looked at the BS transformation on dual FS inputs to a fiber-coupler BS, including scattering losses (due to sidewall roughness).
The authors reported: ``\tit{If the same number state enters both ports of the coupler, the probability of finding an odd number of photons at either of the output ports vanishes for a particular choice of the coupler length}." The authors did not explicitly mention that this length is the one appropriate for a 50:50 fiber-coupler BS, which is most certainly the case. 
Additionally, the authors also considered a FS/CS input including fiber losses, but were primarily concerned with this case being a generator of displaced FS states, with the coupler behaving essentially as a homodyne detector.
The inclusion of photon loss was obtained by averaging the joint photon number distribution over a Bernoulli distribution with a Beer's law loss parameter given by $\nu=e^{-2\gamma\,L}$, resulting again in marginal distributions remaining binomial in the presence of dissipation.
Since the inclusion of loss is essentially straightforward following LBK \cite{Lai:1991} (see Eq.(41-43)), 
and recently Laiho \tit{et al.} \cite{Laiho:2019},
we shall ignore dissipation in the main exposition of this work and consider only an ideal lossless BS. 
The reason for this is so that we can concentrate on the universality of the CNL and PNC features discussed in this work. 
In a later section, we show plots of the output joint photon number distribution from the BS with loss included.
In addition, with today's advances in number-resolving photon detectors (discussed in the penultimate section) 
photon numbers up to $5-6$ can be reliably experimentally distinguished (and at some wavelengths, up to $10-20$, 
see \cite{Gerrits:2016:v2, Schmidt:2018}). We shall return to these considerations at the end of this work.

This paper is organized as follows:
In \Sec{sec:General} we derive our main results concerning the CNL and PNC for a general input states and a BS with an arbitrary transmission $T = \cos^2(\theta/2)$ and reflection coefficient $R = \sin^2(\theta/2)$. 
Based solely on the properties of the BS rotation coefficients, we derive the general condition for the appearance of the CNL and PNC for $a$-mode input states containing only odd numbered (non-classical) FS.
In \Sec{sec:50:50:BS}  we specialize to the important case of a 50:50 BS ($\theta=\pi/2$) and explore the PNC 
in general, as well as for the specific case considered by BMG of a FS/CS input 
state $\ket{n}_a\ket{\beta}_b$ with $n\in\{2,3\}$ 
and an arbitrary CS  of mean photon number $\bar{n}_b = |\beta|^2$. We present analytic formulas for the zeros of the PNC. 
We present illustrative cases for various odd parity $a$-mode input states including odd FS, photon-added single-mode squeezed state, and an odd (CS-based) cat state mixing on a 50:50 BS with a $b$-mode CS $\ket{\beta}_b$.
Additionally, we investigate the case of a non-50:50 BS $\theta=\pi/3$ and explore the existence of analytic solutions for the PNC of the  BMG scenario of an FS/CS input state $\ket{n}_a\ket{\beta}_b$ with $n\in\{2,3\}$ mixing with a  CS.
In \Sec{sec:Experimental} we discuss the possible experimental realization of these effects in light 
of current photon-number resolving transition edge sensors (TES) in the presence of a small amount of noise.
We show plots of the output joint photon number distribution from the BS with and without loss.
Finally, in \Sec{sec:Conclusion} we conclude and discuss our general results. 
In \App{app:BS} we provide the details of the derivation of the specific form of the beam splitter coefficients 
upon which our main results rely.
In \App{app:Proof} we provide the proof of a crucial property of the beam splitter coefficients 
that is central to the appearance of the CNL.
\App{tables:ma:mb:of:k} provides tables of analytic results for zeros indicating complete destructive interference in the case of a non-50:50 beam splitter.
In \App{app:HOM:CNL:Atoms} we discuss the extended HOM effect and the analogue of the CNL for a collection of atoms (Dicke states) using the Schwinger representation of $su(2)$ in terms of pseudo-angular momentum states formed from
two bosonic modes.

\section{General Considerations}\label{sec:General}
The HOM effect \cite{HOM:1987} is the textbook example \cite{Agarwal:2013,Gerry_Knight:2004} of the destructive interference of quantum amplitudes at a BS. 
We choose the convention for the BS unitary transformation (see \App{app:BS})
with transmission coefficient $T=\cos^2(\theta/2)$
as (see \Fig{fig:BS:v1:app})
%
%
\bea{BS:defn:main}
\vec{a}^\dag(\theta) &=& 
\left[
\begin{array}{c}
a^\dag(\theta) \\
b^\dag(\theta)
\end{array}
\right] 
\equiv
U(\theta)
\left[
\begin{array}{c}
a^\dag \\
b^\dag
\end{array}
\right] 
U^\dag(\theta)\no
&=&
\left[
\begin{array}{cc}
\cos(\theta/2) & \sin(\theta/2)\\
-\sin(\theta/2) & \cos(\theta/2)
\end{array}
\right]
\,
\left[
\begin{array}{c}
a^\dag \\
b^\dag
\end{array}
\right] 
\equiv S_{BS}(\theta)\, \vec{a}^\dag, \qquad
\eea
for $0\le\theta\le\pi$, such that a 50:50 BS is given by $\theta = \pi/2$ ($T=R=1/2$).

Consider two single-mode FS $\ket{\Psi_{in}}_{ab} = \ket{1}_a\ket{1}_b = a^\dag b^\dag\ket{0,0}_{ab}$ entering the BS. 
The well known transformation proceeds as follows:
%
\begin{alignat}{2}
& \ket{\Psi_{out}}_{ab} &&= U(\theta)\, a^\dag b^\dag\ket{0,0}_{ab} =a^\dag(\theta)\,b^\dag(\theta)\ket{0,0}_{ab}, \no
%
%
%
&= 
\big[a^\dag\cos&&(\theta/2)+b^\dag \sin(\theta/2) \big] \,
   \left[b^\dag\cos(\theta/2)-a^\dag \sin(\theta/2) \right]\,\ket{0,0}_{ab},  \no
&=
\hspace{1em}
\Big[
\frac{1}{\sqrt{2}} 
&&\left(-\frac{(a^\dag)^2}{\sqrt{2}} + \frac{(b^\dag)^2}{\sqrt{2}}\right)\, \sin(\theta)
+ a^\dag b^\dag \cos(\theta)
\Big]\,
\ket{0,0}_{ab}, \no
&\overset{\theta\to\pi/2}{\longrightarrow}&&
\frac{1}{\sqrt{2}} \left(-\ket{2,0}_{ab} + \ket{0,2}_{ab} \right), \label{HOM:calc}
\end{alignat}
where the minus sign in front of  $\ket{2,0}_{ab}$ arises from our choice of the BS transformation \Eq{BS:defn:main}
and could be removed by a trivial change in phase accomplished by $a^\dag \to i\,a^\dag$ (see Agarwal, Chapter 5 \cite{Agarwal:2013}). The interference of the amplitudes indicated by the $\cos(\theta)$ in front of $\ket{1,1}_{ab}$ arises of course, from the destructive interference of the two possible paths to obtain the coincident measurement $P_{ab}(1,1) = \cos^2(\theta)$, i.e. 
the $a$ and $b$ mode  are either both transmitted, or both reflected into separate detectors, with equal  amplitude magnitudes
$\cos^2(\theta/2)=\sin^2(\theta/2)\overset{\theta\to\pi/2}{\longrightarrow}=1/2$, though opposite in sign, for a 50:50 BS.

In general, 
since any two-mode state (pure or mixed) can be expanded in a dual FS number basis $\ket{n,m}_{ab}$ we need to know how each basis state transforms under the action of the BS.
As shown in \App{app:subsec:BS:derivation} (see also Agarwal, \tit{Quantum Optics}, Chapter 5 \cite{Agarwal:2013}, LBK \cite{Lai:1991}) we have
\bwt
\bsub
\bea{nm:under:BS:main}
\hspace{-0.25in}
\ket{n,m}^{(out)}_{ab} &=& U(\theta) \ket{n,m}_{ab}, \no
&=&  \sum_{p=0}^{n+m}\, f^{(n,m)}_p(\theta) \,\ket{p}_a\,\ket{n+m-p}_b, \label{nm:under:BS:line1:main} \\
\hspace{-0.25in}
f^{(n,m)}_p(\theta) &=&
 \sum_{q=0}^n  \sum_{q'=0}^m \delta_{p,q+q'}\, 
 \binom{n}{q}\,\binom{m}{q'}\,
\sqrt{\dfrac{p!\,(n+m-p)!}{n!\,m!}}\,
\left(-1\right)^{q'}\,
   \left(\cos(\theta/2)\right)^{m+(q-q')}\,
   \left(\sin(\theta/2)\right)^{n-(q-q')}, \qquad \label{nm:under:BS:line2:main} \\
&=&   
\sqrt{\dfrac{p!\,(n+m-p)!}{n!\,m!}}\,
 \left(\cos(\theta/2)\right)^{m-p}
   \left(\sin(\theta/2)\right)^{n+p}\,
   \left(-1\right)^{p}\,
 \sum_{q=0}^n \binom{n}{q}\,\binom{m}{p-q}\,\left(\frac{-1}{\tan^2(\theta/2)}\right)^q.  \label{nm:under:BS:line3:main} 
\eea
\esub
\ewt
%
In \App{app:BS}, we show that through clever use of the lower factorial function 
$(x)_k \equiv (x-0)(x-1)(x-2)\cdots(x-(k-1)) = \tfrac{x!}{(x-k)!}$
\cite{Pochhammer:comment}
 we can rewrite \Eq{nm:under:BS:line3:main} in the compact form 
 (see \Eq{fnmp:final:line1} and \Eq{gnmptheta})
\bwt
\bsub
\bea{fnmp:Final:main}
f^{(n,m)}_p(\theta) 
&\equiv& 
\frac{(-1)^{p+n}}{\sqrt{n!\, (m+n)_n}}\,\sqrt{\binom{m+n}{p}}\, 
\big(\cos(\theta/2) \big)^{m-p}\,\big(\sin(\theta/2) \big)^{p-n}\,g^{(n,m)}_p(\theta), \label{fnmp:Final:line1:main} \\
g^{(n,m)}_p(\theta) &\equiv& \sum_{q=0}^{n} \binom{n}{q}\,(-1)^q\, (p)_{n-q}\,
 \big(\cos(\theta/2) \big)^{n-q} \,(m+n-p)_{q}\,\big(\sin(\theta/2) \big)^{q}. \label{fnmp:Final:line2:main}
\eea
\esub
\ewt

Of primary interest will be the probability amplitude $\mathcal{A}(m_a, m_b,\theta|n,m)$
 for the measurement (projection)  $M_{ab}\equiv\ket{m_a, m_b}_{ab}\bra{m_a, m_b}$
of $m_a$ photons in mode $a$ and $m_b$ photons in mode $b$ given by 
\bsub
\bea{Amamb:nm:main}
\mathcal{A}(m_a, m_b,\theta|n,m) &=&  {}_{ab}\IP{m_a,m_b}{n,m}^{(out)}_{ab}, \label{Amamb:nm:line1:main} \\
&=&  \sum_{p=0}^{n+m}\, f^{(n,m)}_p(\theta) \, _a\langle m_a\ket{p}_a\,   _b\langle m_b\ket{n+m-p}_b, \no
&=& f^{(n, m_a+m_b-n)}_{m_a}(\theta) \, \delta_{m,m_a+m_b-n}. \label{Amamb:nm:line2:main}
\eea
\esub
The delta function $\delta_{m+n, m_a+m_b}$ ensures that one can only measure output states $\ket{m_a,m_b}$ 
exiting the BS 
such that the output sum $m_a+m_b$ is equal to the total number of photons $n+m$ entering the BS.
The particular machinations that we have made in \Eq{fnmp:Final:line2:main} were in the anticipation 
of measurement $M_{ab}$ that 
induces $p\to m_a$ and $m+n-p\to (m_a+m_b-n) + n - m_a = m_b$, as given by the delta function in \Eq{Amamb:nm:line2:main}.
Using \Eq{fnmp:Final:line1:main} and \Eq{fnmp:Final:line2:main} we explicitly have (see \Eq{fmambthetan:line1} and \Eq{fmambthetan:line2})
\bwt
\bsub
\bea{fmambthetan:Final}
\hspace{-0.5in}
f^{(n,m_a+m_b-n)}_{m_a}(\theta) &=& 
\frac{(-1)^{m_a+n}}{\sqrt{n!\, (m_a+m_b)_n}}\,\sqrt{\binom{m_a+m_b}{m_a}}\, 
\big(\cos(\theta/2) \big)^{m_b-n}\,\big(\sin(\theta/2) \big)^{m_a-n}\,g^{(n,m_a+m_b-n)}_p(\theta), \label{fmambthetan:Final:line1} \qquad\\
\hspace{-0.5in}
g^{(n,m_a+m_b-n)}_{m_a}(\theta) &=&
\delta_{m_a+m_b,n+m}\,
\sum_{q=0}^{n} \binom{n}{q}\,(-1)^q\, (m_a)_{n-q}\, \big(\cos^2(\theta/2) \big)^{n-q} \,(m_b)_{q}\,\big(\sin^2(\theta/2) \big)^{q}
\equiv g(m_a,m_b|n,m). \qquad\quad
\label{fmambthetan:Final:line2}
\eea
\esub
\ewt
The implications of 
 \Eq{fmambthetan:Final:line1} and  \Eq{fmambthetan:Final:line2} are the main focus of this work.

Let us now consider a general bipartite (in general, mixed) state $\rho_{ab}$ impinging on the BS with input modes $a$ and $b$.
The probability $P(m_a,m_b,\theta)$ to measure $m_a$ photons in the output mode $a$ and $m_b$ photons in the output mode $b$ upon exit from the BS is given by
\bwt
\bsub
\hspace{-0.35in}
\bea{BS:on:rho:ab}
\hspace{-0.35in}
&{}& \rho_{ab} = \sum_{n,n',m,m'} \rho_{n m, n' m'} \ket{n,m}_{ab}\,{}_{ab}\bra{n',m'}, \label{BS:on:rho:ab:line1} \no
&\overset{U(\theta)}{\longrightarrow}& \rho^{(out)}_{ab} = 
 \sum_{n,n',m,m'} \rho_{n m, n' m'} \ket{n,m}^{(out)}_{ab} {}_{ab}^{(out)}\bra{n',m'},\\
\hspace{-0.35in}
&\Rightarrow&  P(m_a,m_b,\theta) = \Tr[M_{ab}\,\rho^{(out)}_{ab}], \label{BS:on:rho:ab:line1} \no
&=& 
 \sum_{n,n',m,m'} \rho_{n m, n' m'}  {}_{ab}\IP{m_a,m_b}{n,m}^{(out)}_{ab}\;
                                                        {}^{(out)}_{ab}\IP{n',m'}{m_a,m_b} {}_{ab}, \no
 \hspace{-0.35in}
 &=& 
  \sum_{n,n',m,m'} \rho_{n m, n' m'} \; f^{(n, m_a+m_b-n)}_{m_a}(\theta) \, \delta_{m,m_a+m_b-n}\;
                                                        f^{(n', m_a+m_b-n')}_{m_a}(\theta) \, \delta_{m',m_a+m_b-n'}.    \qquad\qquad  \label{BS:on:rho:ab:line2}                                              
\eea
\esub
\ewt
Let us designate pure states in mode $a$ and $b$ as 
$\ket{\psi}_a = \sum_{n=0}^\infty c_{n} \ket{n}_a$ 
and 
$\ket{\phi}_b = \sum_{m=0}^\infty d_{m} \ket{m}_b$
with 
$\sum_{n=0}^\infty |c_{n}|^2= \sum_{m=0}^\infty |d_{m}|^2=1$,
and 
a mixed state $\rho_b$ in mode $b$ as $\sum_{m,m'=0}^\infty \rho^{(b)}_{m,m'} \ket{m}_b\bra{m'}$ 
with $ \sum_{m=0}^\infty  \rho^{(b)}_{m,m}=1$. 
Specializing the general case \Eq{BS:on:rho:ab:line2}  to interesting specific cases we obtain
\bwt
\bsub
\begin{alignat}{4}
%
%
&\trm{FS/FS:}\quad \ket{\Psi_{in}}_{ab} &&= \ket{n,m}_{ab} 
&&\Rightarrow 
P(m_a,m_b,\theta) &&= \Big|f^{(n, m_a+m_b-n)}_{m_a}(\theta)\Big|^2,  
\label{P:for:input:states:FS:FS} \\
%
%
&\trm{FS/PS:}\quad \ket{\Psi_{in}}_{ab} &&= \ket{n,\phi}_{ab} 
&&\Rightarrow 
P(m_a,m_b,\theta) &&= \big| \, d_{m_a+m_b-n}\big|^2 \, \big| f^{(n, m_a+m_b-n)}_{m_a}(\theta)\big|^2, 
\label{P:for:input:states:FS:PS} \\
%
%
&\trm{FS/MS:}\quad \rho^{(in)}_{ab} &&= \ket{n}_{a}\bra{n}\otimes\rho_b
&&\Rightarrow
P(m_a,m_b,\theta) &&= \rho^{(b)}_{m_a+m_b-n, m_a+m_b-n} \,\big| f^{(n, m_a+m_b-n)}_{m_a}(\theta)\big|^2, \label{P:for:input:states:FS:MS} \\  
%
%
&\trm{PS/PS:}\quad \ket{\Psi_{in}}_{ab} &&= \ket{\psi,\phi}_{ab} 
&&\Rightarrow
P(m_a,m_b,\theta) &&= \Big|\sum_{n=0}^\infty c_n \, d_{m_a+m_b-n} \,f^{(n, m_a+m_b-n)}_{m_a}(\theta)\Big|^2,  \label{P:for:input:states:PS:PS} \\
%
%
&\trm{PS/MS:}\quad \ket{\Psi_{in}}_{ab} 
&&= \ket{\psi}_a\bra{\psi}\otimes\rho_b 
&& \Rightarrow 
P(m_a,m_b,\theta) \nonumber 
\end{alignat}
\be{P:for:input:states:PS:MS}
\hspace{1.0in}\quad =  
\left|
\sum_{n,n'=0}^\infty
c_n\,c^*_{n'}\,
\rho^{(b)}_{m_a+m_b-n, m_a+m_b-n'} \, f^{(n, m_a+m_b-n)}_{m_a}(\theta) \,f^{(n', m_a+m_b-n')}_{m_a}(\theta) 
\right|^2,
\ee
\esub
\ewt
where we have used the abbreviations: FS: Fock state, PS: pure state, MS: mixed state, and from now on
drop the subscript $ab$ on the output joint probability distribution $P(m_a,m_b,\theta)$.

The important point to note is that for 
\Eq{P:for:input:states:FS:FS}, \Eq{P:for:input:states:FS:PS} and \Eq{P:for:input:states:FS:MS} with a
single FS $\ket{n}_a$ in the $a$ mode, we isolate a \tit{single} beam splitter coefficient (BSC)
$f^{(n, m_a+m_b-n)}_{m_a}(\theta) \propto g(m_a,m_b,\theta)$, \Eq{fmambthetan:Final:line2}. 
Therefore, if 
$g(m_a,m_b,\theta)\to 0$, then the joint probability 
$P(m_a,m_b,\theta)\to 0$ \tit{independent} of the amplitudes/density matrix elements that define the $b$-mode input state.
We shall see shortly that this always occurs for the case of $n$ \tit{odd} 
(where $n$ labels the FS of the $a$-mode input state)
and for a 50:50 BS ($\theta=\pi/2$) when $m_a=m_b=m'$, i.e.
$P(m',m',\pi/2)=0$ for
joint coincidences of the same number of photons in the output modes of the 50:50 BS.

For the situations of a pure state in the $a$ mode and either pure or mixed state in the $b$ mode, from 
 \Eq{P:for:input:states:PS:PS} and \Eq{P:for:input:states:PS:MS} we see that we have additional sums over the amplitudes and/or density matrix elements defining the $a$- and $b$-mode input states. 
 However, if the $a$-mode state contains \tit{only} a superposition of \tit{odd} number of photons (FS), 
 i.e. an odd parity state,
 then when each of the BSCs is evaluated at $\theta=\pi/2$ for a 50:50 configuration \tit{each} of the BSC will separately go to zero, independent of the amplitudes/density matrix elements defining the states, and once again the diagonal of the joint probability $P(m',m',\pi/2)$, will go to zero.
The proof of these claims is developed in the next section.

\section{Angular portion $\boldsymbol{g(m_a,m_b,\theta)}$ for a 50:50 BS}\label{sec:50:50:BS}
The focus of this section is to explore the angular portion $g(m_a,m_b,\theta)$,  \Eq{fmambthetan:Final:line2},
of the probability amplitude ${}_{ab}\IP{m_a,m_b}{n,m}^{(out)}_{ab}$ arising from the BS amplitude
 $f^{(n,m)}_p(\theta)$ \Eq{fnmp:Final:line1:main} 
of the transformed basis state $\ket{n,m}^{(out)}_{ab} = U(\theta) \ket{n,m}_{ab}$
after the measurement $M_{ab}$ is made of $m_a$, $m_b$ photons in the outut $a$ and $b$ modes, respectively.
Of particular interest will be $g(m_a,m_b,\pi/2)$ in the 50:50 BS configuration $\theta=\pi/2$.

\subsection{$\boldsymbol{g(m_a,m_b,\theta)}$: the general case}
In general, $g(m_a,m_b,\theta)$  is given by 
\bea{gmambn:general}
\lefteqn{g(m_a,m_b,\theta|n, m) = 
\delta_{m_a+m_b,n+m}\times}\, \no
&{ }& 
\sum_{q=0}^{n} \binom{n}{q} \,(-1)^q \,
(m_a)_{n-q}\,\left(\cos^2(\theta/2)\right)^{n-q}\,
(m_b)_{q}\,\left(\sin^2(\theta/2)\right)^{q}, \qquad
\eea
where $(x)_q$ is the \tit{lower or descending factorial function} 
given by
\bea{lower:factorial:function}
\hspace{-0.05in}
(x)_q \equiv x_{(q)} &=& (x-0)(x-1)(x-2)\cdots(x-(q-1)) =  \frac{x!}{(x-q)!}\no
  (x)_0&\equiv& 1,\;\;    (x)_1=x,\;\;   (x)_2 = x(x-1),\ldots. \qquad
\eea
\Eq{gmambn:general} generalizes Eq.(7-10) of the angular portion of the formulas in BMG \cite{BMG:2012}.
For $n\in\{1,2,3\}$ \Eq{gmambn:general}  explicitly yields
\bwt
\bsub
\bea{BMG:nodal:lines:theta}
g(m_a,m_b,\theta|1) &=& \left(m_a\cos^2(\theta/2)-m_b\sin^2(\theta/2)\right), \label{BMG:nodal:lines:theta:n1} \\
%
%
g(m_a,m_b,\theta|2) &=& \left(m_a\cos^2(\theta/2)(m_a-1)\cos^2(\theta/2)-2m_a\cos^2(\theta/2) m_b\sin^2(\theta/2) + m_b\sin^2(\theta/2)(m_b-1)\sin^2(\theta/2)\right)), \qquad \label{BMG:nodal:lines:theta:n2} \\
\hspace{-0.75in}
g(m_a,m_b,\theta|3) &=& \left(m_a\cos^2(\theta/2)(m_a-1)\cos^2(\theta/2)(m_a-2)\cos^2(\theta/2)-3m_a\cos^2(\theta/2)(m_a-1)\cos^2(\theta/2)m_b\sin^2(\theta/2)\right. \no
%
%
&+& \left.3m_a\cos^2(\theta/2) m_b\sin^2(\theta/2)(m_b-1)\sin^2(\theta/2)-m_b\sin^2(\theta/2)(m_b-1)\sin^2(\theta/2)(m_b-2)\sin^2(\theta/2)\right)),\qquad \quad \label{BMG:nodal:lines:theta:n3}
\eea
\esub
\ewt
where $P(m_a, m_b,\theta|n)\propto g^2(m_a,m_b,\theta|n)$.
Here we have written $g(m_a,m_b,\theta|n)$ and $P(m_a,m_b,\theta|n)$ since the value of $m$ is given by the delta function
$\delta_{m_a+m_b,n+m}$ in \Eq{gmambn:general} that ensures that the total number of photons $m_a+m_b$ measured upon exit of the BS is equal to the total number of input photons $n,m$ from the dual basis state  $\ket{n,m}_{ab}$.

We see that in \Eq{gmambn:general} and explicitly in \Eq{BMG:nodal:lines:theta:n1}-\Eq{BMG:nodal:lines:theta:n3} that each factor of $(m_a-j)_{j\in\{0,\ldots((n-k)-1)\}}$ in $(m_a)_{n-k}$ is associated with a factor of $T=\cos^2(\theta/2)$
and similarly that each factor of
$(m_b-j)_{j\in\{0,\ldots(k-1)\}}$ in $(m_b)_{k}$ is associated with a factor of $R=\sin^2(\theta/2)$.

Note that for a single $a$-mode input FS $\ket{1}_a$,  \Eq{BMG:nodal:lines:theta:n1} indicates that  
individual nodes or zeros of  $P(m_a, m_b,\theta|1)$ are given 
for a general BS angle $\theta$ by
$\tan^2(\theta/2) = m_a/m_b$, and \tit{not} just for the 50:50 BS case of BMG of $\theta=\pi/2 \Rightarrow m_a=m_b$.
However, we are ultimately interested in $\tit{nodal curves}$, i.e. \tit{curves of zeros} representing complete destructive interference that occur for some relationship between $m_a$ and $m_b$ and for some angle $\theta$. Our primary result is that we can always produce such a nodal line 
\tit{for a 50:50 BS ($\theta=\pi/2$) with $m_a=m_b=m'$} if the input state to mode $a$ is an \tit{odd numbered FS}
$\ket{n=2 n'+1}_a$ \tit{regardless} of the input state entering the $b$ mode.

\Eq{gmambn:general} has a kind of  \tit{hyper-binomial formula} structure.
The function
$g(m_a,m_b,\theta|n)$ can be formally summed as
\bea{gmambn:Mathematica}
\hspace{-0.25in}
\lefteqn{
g(m_a,m_b,\theta|n,m) = \delta_{m_a+m_b,n+m}\, (m_a)_n 
} \no
&\times&
\left(\cos^2(\theta/2)\right)^{n}\,  _{2}F_{1}(-m_b,-n; 1+m_a-n; -\tan^2(\theta/2)),\quad\;\;\;
\eea
where $_{2}F_{1}(a, b; c; z) = \sum_{k=0}^{\infty} \tfrac{ (a)^{(k)}\,(b)^{(k)}}{(c)^{(k)}}\, \tfrac{z^k}{k!}$
is the hypergeometric function with
$x^{(n)} = (x+0)(x+1)(x+2)\cdots(x+(n-1)) = \tfrac{\Gamma(z+k)}{\Gamma(z)}$, 
$x^{(0)} \equiv 1$, $x^{(1)} = x$, $x^{(2)} = x(x+1)$,\ldots is the
\tit{rising or ascending factorial function} or \tit{Pochhammer symbol}
\cite{Pochhammer:comment}.
While \Eq{gmambn:Mathematica} reproduces \Eq{gmambn:general}, the former involves ratios of Gamma functions of negative integers, which individually are infinite, but whose ratios are finite. 
However, we find \Eq{gmambn:general} more intuitively appealing and useful due to its \tit{hyper-binomial formula} structure.

\subsection{Effects at a 50:50 BS: CNL for $\boldsymbol{n}$ odd, $\boldsymbol{m_a=m_b= m}$}\label{subsec:50:50:BS}
An examination of \Eq{gmambn:general} reveals that it is invariant, up to a crucial overall sign $(-1)^n$, under the interchange of 
$m_a\leftrightarrow m_b$ and $\theta\to \pi-\theta$, i.e. 
\be{g:properties}
g(m_b,m_a, \pi-\theta|n) = (-1)^n\, g(m_a,m_b,\theta|n).
\ee
This can be seen by noting that $\theta\to \pi-\theta$ exchanges $\cos(\theta/2) \leftrightarrow \sin(\theta/2)$.
When setting $m_a\leftrightarrow m_b$ in \Eq{gmambn:general} 
we note that we can write the binomial coefficient $\binom{n}{q}$ as
$\binom{n}{n-q}$. Defining $q'=n-q$ or $q=n-q'$ interchanges the indices $q$ and $n-q$ to $n-q'$ and $q'$, respectively.
Finally, the term $(-1)^q\to (-1)^{n-q'} =(-1)^{n}\, (-1)^{q'}$. Therefore, upon factoring out the term $(-1)^{n}$
we can relabel $q'\to q$ to finally arrive at \Eq{g:properties}.

Thus, for an input state $\ket{n,m}_{ab}$, and
for the case of joint coincident counts with equal number of measured photons in each output port of a 50:50 BS,
 i.e. ($\theta=\pi/2$) and $m_a = m_b \equiv m'$,  \Eq{g:properties} yields
\bea{g:mambeqm:50:50:BS}
\hspace{-0.05in}
\ket{\Psi_{in}}_{ab} &=& \ket{n,m}_{ab}, \;\trm{and}\;\;
\trm{50:50 BS}\,(\theta = \pi/2), \;
\trm{and}\; m_a=m_b\equiv m': \no
 &\Rightarrow& \;
g(m',m', \pi/2|n) = (-1)^n\, g(m',m',\pi/2|n) 
\overset{n\,\trm{odd}}{\longrightarrow} 0, \no
 &\Rightarrow & P(m',m', \pi/2|n)\overset{n \trm{odd} }{=}0.
\eea
That is, for $n$ \tit{odd} we have a \tit{central nodal line} (CNL) of contiguous zeros \tit{for all} integers $m\in\mathbb{Z}_{\ge0}$ in the $b$-mode. 
This result depends \tit{solely} on the intrinsic properties of the BS in a 50:50 configuration and is independent of the $b$-mode  input FS state, as long as $n$ is odd for the $a$-mode input Fock state.

The above result is evident in \Eq{BMG:nodal:lines:theta:n1}  where for $\theta=\pi/2$ and 
$n=1$ we have 
\bsub
 \be{gmm:cancellation:n1}
g(m_a,m_b,\pi/2|1) = 
\frac{1}{4}\,(m_a-m_b)
\overset{m_a=m_b}{\longrightarrow} 0,
\ee
and for $n=3$ from  \Eq{BMG:nodal:lines:theta:n3} we have
\bea{gmm:cancellation:n3}
g(m_a,m_b,\pi/2|3) &=& 
\frac{1}{8}\,
\Big((m_a)_3 - 3\, (m_a)_2\,(m_b)_1 + \no
&+&
3\,(m_a)_1(m_b)_2 - (m_b)_3 \Big) 
\overset{m_a=m_b}{\longrightarrow} 0. \qquad\quad 
\eea
\esub
In general, for $n$ \tit{odd} there are $n+1\to$ even number of terms in $g(m_a,m_b,\pi/2|n)$ 
with a symmetric binomial coefficient $\binom{n}{q}$ with alternating signs $(-1)^q$ so that  the first $(n+1)/2$ terms have equal magnitude, but opposite signs of the later $(n+1)/2$ terms (e.g.  for $n=5$, the alternating binomial coefficients are
$(1,-5,10,-10,5,-1)$). When $m_a=m_b\to m$ as for a 50:50 BS, these even number of terms cancel in pairs to yield zero. 

For $n=1$,  $g(m',m', \theta|n=1) = m'\,\cos(\theta)\overset{\theta=\pi/2}{\longrightarrow} 0$ for a 50:50 BS.
A closer examination of 
\Eq{gmambn:general}
and
\Eq{gmambn:Mathematica} 
reveals 
the following behavior for $g(m',m', \theta|n)$ for $n$ odd and even:
\bsub
\bea{gmmtheta:n:odd}
\hspace{-0.15in}
&{}&n = 2n'+1\; \trm{odd:} \\
\hspace{-0.15in}
&{}&g(m',m', \theta|n=2n'+1)
=\tfrac{(m')_{n'}}{2^{3\,n'-1}}\,\cos(\theta)\,\trm{poly}_{n'}(m',\theta)
\overset{\theta=\pi/2}{\longrightarrow} 0,  \no
\hspace{-0.15in}
&{}& \no
\hspace{-0.15in}
\label{gmmtheta:n:even} 
&{}&n = 2n'\; \trm{even:}\\
\hspace{-0.15in}
&{}&g(m',m', \theta|n=2n')
=\tfrac{(m')_{n'+1}}{2^{3\,n'-1}}\,\trm{poly}^{\prime}_{n'}(m',\theta)
\overset{\theta=\pi/2}{\ne} 0. \nonumber 
\eea
\esub
In \Eq{gmmtheta:n:odd} with $n'>0$,
$\trm{poly}_{n'}(m',\theta)$ is a polynomial of order $n'$ in $m$ and involves $\cos(k\,2\,\theta)$ terms for 
$k\in\{1,2,\ldots n'\}$ such that  $\trm{poly}_{n'}(m',\theta=\pi/2)\ne 0$.
The \tit{crucial} point is that for $n$  \tit{odd}  a factor of $\cos(\theta)$ 
can \tit{always} be factored out of the full expression for
$g(m',m', \theta|n)$ (see proof in \App{app:Proof}), 
with $\cos(\theta=\pi/2)=0$ for a 50:50 BS.
%
For $n$ even, \tit{no} such trigonometric terms factors out of the full expression for $g(m',m', \theta|n)$ in \Eq{gmmtheta:n:even}.
In \Eq{gmmtheta:n:even} $\trm{poly}^{\prime}_{n'}(m,\theta)$ has the same characteristics (form) as $\trm{poly}_{n'}(m,\theta)$ in \Eq{gmmtheta:n:odd}, 
most importantly that $\trm{poly}^{\prime}_{n'}(m,\theta=\pi/2)\ne 0$ for a 50:50 BS.

It should be noted that half the points on the CNL for the dual FS input $\ket{n,m}^{(in)}_{ab}$
have $P(m',m', \pi/2|n=odd) =0$ \tit{trivially}, since if $m=even$ there simply is no output state
$\ket{m',m'}^{(out)}_{ab}$ with equal number of photons both the $a$ and $b$ modes.
For example for input state $\ket{1,2}^{(in)}_{ab}$, the output state $\ket{1,2}^{(out)}_{ab}$ from a 50:50 BS
 is spanned by the basis states 
$\{\ket{0,3}_{ab}, \ket{1,2}_{ab}, \ket{2,1}_{ab}, \ket{3,0}_{ab}\}$.
Thus, no complete destructive interference is occurring (the HOM effect) since a diagonal output state is simply not present.
However, if $m$ is \tit{also odd}, then $n+m$ is even and the output contains 
the diagonal state $\ket{m',m'}_{ab}$ with $m'=(n+m)/2$ 
\tit{whose coefficient is proportional to $\cos(\theta)\overset{\theta=\pi/2}{\longrightarrow} 0$}. This \tit{is} the HOM effect, i.e. complete destructive interference for the amplitude of the diagonal output state of a 50:50 BS.
%
%
In a sense, the 50:50 BS acts like a ``(notch) filter" that singles out FS output pairs $\ket{m',m'}_{ab}$ (with $n, m$ both odd) and ``tag" them with a factor proportional to $\cos(\theta)$, which can then be made zero for a 50:50 BS.
For example, for input state $\ket{1,3}^{(in)}_{ab}$ the output states are spanned by 
$\{\ket{0,4}_{ab}, \ket{1,3}_{ab}, \ket{2,2}_{ab}, \ket{3,1}_{ab}, \ket{4,0}_{ab}\}$, and the diagonal output state $\ket{2,2}_{ab}$ has a coefficient proportional to $\cos(\theta)$ for a BS of arbitrary reflectivity.
Note that the input state $\ket{0,4}^{(in)}_{ab}$ (as well as input states $\ket{2,2}^{(in)}_{ab}$ and 
$\ket{4,0}^{(in)}_{ab}$) also contain the diagonal output state $\ket{2,2}_{ab}$. 
However, because $n$ (and $m$) is even in this case one will \tit{not} be able to factor out the term $\cos(\theta)$ from $f_{p=2}^{n=2, m=2}(\theta)$, and hence \tit{no} HOM effect occurs.

Explicitly, for input states $\ket{2,m}$ there are no output states of the form $\ket{m',m'}_{ab}$ with $m$ odd since that would imply that $m'=(n+m)/2$ would be a half-integer. For $m=even$ we have 
from \Eq{gmmtheta:n:even}  that 
$P(m',m',\theta|n=2) = \tfrac{1}{4}\,m'\,\big( 2m'-3 + (2m'-1)\,\cos(2\theta)\big)$
for $m'=(n+m)/2$. Thus, input states $\{\ket{2,2}_{ab}, \ket{2,4}_{ab}, \ket{2,6}_{ab},\ldots\}$ contain output states
$\{\ket{2,2}_{ab}, \ket{3,3}_{ab}, \ket{4,4}_{ab},\ldots\}$ respectively, 
but their coefficients are \tit{not} proportional to $\cos(\theta)$.
If we instead consider input states $\ket{3,m}_{ab}$, states 
with $m=even$ again simply do not produce diagonal output states.
The remaining input states  $\{\ket{3,1}_{ab}, \ket{3,3}_{ab}, \ket{3,5}_{ab},\ldots\}$ wtih $m=odd$ do produce diagonal ouput states
$\{\ket{2,2}_{ab}, \ket{3,3}_{ab}, \ket{4,4}_{ab},\ldots\}$ respectively. From  \Eq{gmmtheta:n:odd} we have
$P(m',m',\theta|n=2) = \tfrac{1}{4}\,m'\,(m'-1)\,\cos(\theta)\,\big( 2m'-7 + (2m'-1)\,\cos(2\theta)\big)$ which is proportional to 
$\cos(\theta)\overset{\theta=\pi/2}{\longrightarrow}0$,  and hence gives rise to the extended HOM effect for a 50:50 BS.
(Note: for input states $\ket{1,m}_{ab}$, we have $P(m',m',\theta|n=1)= m'\,cos(\theta)$ with $m'=(1+m)/2$ which again implies that the extended HOM effect occurs for $n=1, m=odd$.)

The main result here is that for the dual FS input states $\ket{n,m}_{ab}$, \tit{with both $n$ and $m$ odd}, 
the diagonal of the output joint probability distribution 
$P(m'=(n+m)/2,m',\theta|n) \propto \cos^2(\theta)\overset{\theta=\pi/2}{\longrightarrow} 0$ for a 50:50 BS, which is an extended form of the HOM effect. These zeros make up half the points on the output diagonal CNL.

\begin{figure*}[ht]
\begin{tabular}{cc}
\includegraphics[width=3.0in,height=2.25in]{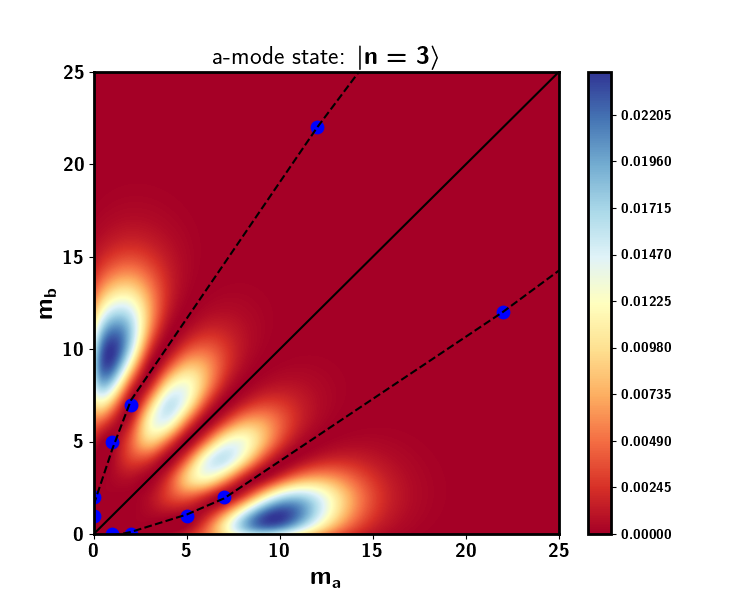} \qquad& 
\includegraphics[width=3.0in,height=2.25in]{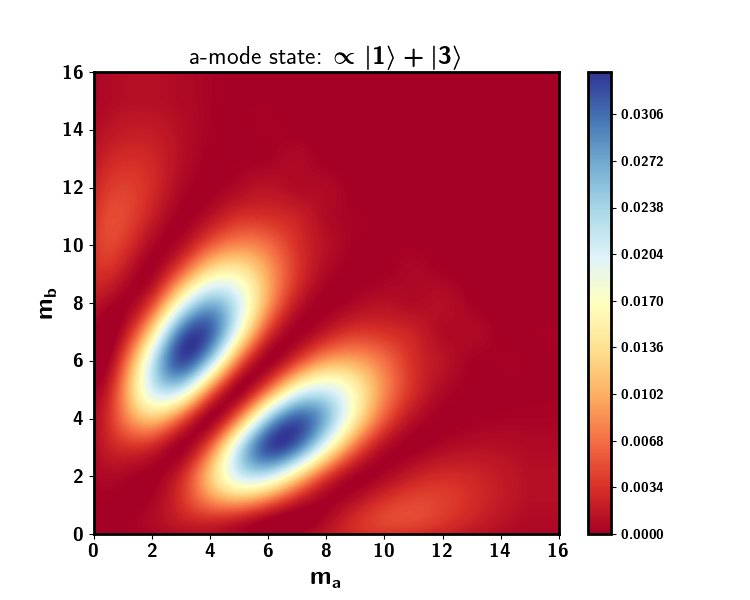} \\
\includegraphics[width=3.0in,height=2.25in]{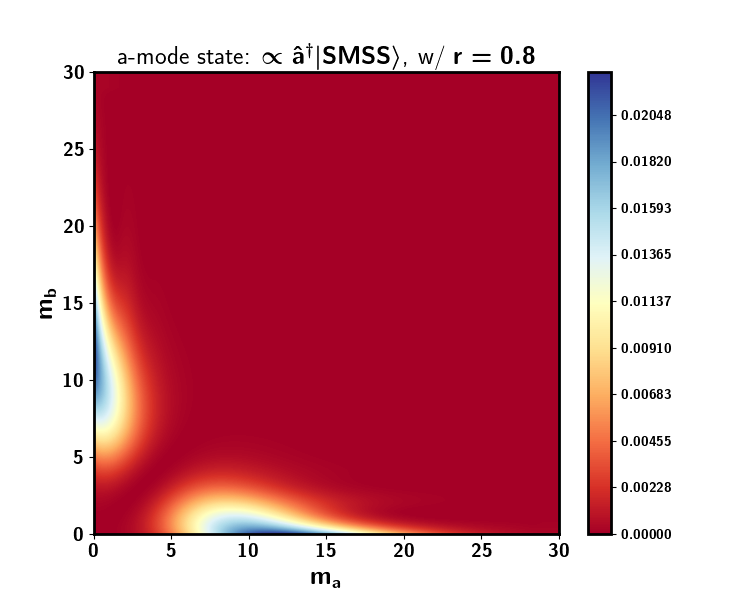} & 
\includegraphics[width=3.0in,height=2.25in]{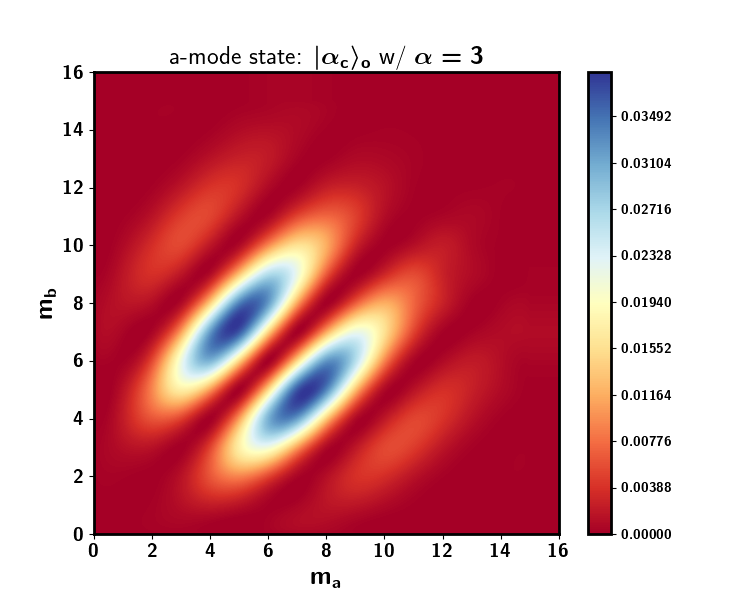} 
\end{tabular}
\caption{
Contour plots of the output joint probability $P(m_a,m_b,\theta=\pi/2)$
for a coherent state $\ket{\beta=3}_b$ entering mode $b$ of a 50:50 BS, 
and $a$-mode input states containing \tit{only odd} numbers of photons:
(top left) $\ket{3}_a$, 
(top right) $(\ket{1}_a+\ket{3}_a)/\sqrt{2}$, 
(bottom left) photon-added single mode squeezed state:   $a^\dag \ket{SMSS}_a$,
(bottom right) odd cate state $\ket{\alpha_c}_a = \mathcal{N}^{-1}\,\left( \ket{\alpha}_a- \ket{-\alpha}_a\right)$,
\tit{all} showing a \tit{central nodal line} CNL, along the diagonal $m_a=m_b$.
The black lines and blue dots in the top left figure for $n=3$ are explained in the text.
(Note: All contour plots in this work are constructed by interpolating discrete data sets
as displayed in the histogram plots of
\Fig{fig:FSCS:FSTS:plots:n0:1}  and \Fig{fig:FSCS:FSTS:plots:n2:3}).
}\label{fig:rb:plots:18Aug2021}
\end{figure*}

The above results apply to each two-mode FS basis input state $\ket{n,m}_{ab}$ entering the BS.
However, since we can expand any $b$-mode pure state $\ket{\psi}_b = \sum_m d_m \ket{m}_b $  in terms of basis states 
$\{\ket{m}_b\}$ the above results also apply to any input state of the form $\ket{n}_a\ket{\psi}_b$
Therefore, \Eq{g:mambeqm:50:50:BS} has wide ranging, universal implications. For $a$-mode input states that contain superpositions of \tit{only odd} FS, the probability $P(m_a,m_b,\theta)$ will involve  the square of a sum of BCS  $g(m_a,m_b,\theta|n\to odd)$ \tit{each of which goes to zero} for a 50:50 BS and for $m_a=m_b$, \tit{regardless of the input state entering mode $b$, pure or mixed}. 
This can be seen from the general formula for the joint photon probability distribution from 
\Eq{P:for:input:states:FS:FS}-\Eq{P:for:input:states:PS:MS} when $n$ (or $n'$) is odd in each 
BSC $f^{(n, m_a+m_b-n)}_{m_a}(\theta)$.
Such candidate example states include (but are not limited to)
$(\ket{1}_a + \ket{3}_a)/\sqrt{2}$, 
a 
photon-added single-mode squeezed state (SMSS), $\cosh^{-1}(r)\,a^\dag \ket{\xi}_a =S(\xi)\ket{1}_a$, 
(with SMSS $\ket{\xi}_a = S(\xi)\ket{0}_a= e^{(\xi a^{2\dag} - \xi^* a^2)/2}\ket{0}_a 
= \tfrac{1}{\sqrt{\cosh (r)}} \sum_{n=0}^\infty e^{i\,n\,\varphi}\tanh^n (r) \tfrac{\sqrt{(2n)!}}{n! 2^n} \ket{2 n}_a$, with 
$\xi = r e^{i\,\varphi}$, (see Agarwal, Chapter 2, (2.18, 2.24), p32, \cite{Agarwal:2013}))
which contains only odd number FS (see Agarwal, Chapter 4.6, p89-90, \cite{Agarwal:2013}), 
or equivalently a
photon-subtracted single-mode squeezed state $ e^{-i\,\varphi} \sinh^{-1}(r)\,a \,\ket{\xi}_a =S(\xi)\ket{1}_a$, 
and 
an  \tit{odd cat} state $\ket{\alpha_c}_a = \mathcal{N}^{-1}\,\left( \ket{\alpha}_a- \ket{-\alpha}_a\right) 
= 2\mathcal{N}^{-1}\left(e^{-|\alpha|^2/2} \sum_{n=0}^\infty \tfrac{s_n\,\alpha^n}{\sqrt{n!}}\right)$ with
$s_n\equiv \tfrac{1 - (-1)^n}{2} = \tiny{\left\{  \begin{tabular}{c}0,\, $n$ even \\ 1,\, $n$ odd \end{tabular} \right.}$, 
and normalization factor 
$\mathcal{N} = (2 - 2 e^{-2 |\alpha|^2} )^{1/2}$ (see Agarwal, Chapter 4.1, p76ff, (4.2) and (4.4) with phase choice $\varphi = \pi$, \cite{Agarwal:2013}).
The joint photon number probability $P(m_a, m_b)$ is plotted below for the above states in \Fig{fig:rb:plots:18Aug2021}, 
 clearly showing a CNL along the diagonal $m_a=m_b$.

The case of an  odd cat state mixed with a  CS is 
a simple, illustrative example of how this CNL occurs for pure states of odd parity
entering mode $a$ and \tit{any} pure state entering  mode $b$, \Eq{P:for:input:states:PS:PS}.
The odd cat state input can be written as 
$
\ket{\Psi_{in}}_{ab} = \ket{\alpha}_a\ket{\beta}_b= 
2\mathcal{N}^{-1}\,e^{-(|\alpha|^2+|\beta|^2)/2} \sum_{n=0}^\infty  \sum_{m=0}^\infty \tfrac{\alpha^n\,s_n}{\sqrt{n!}}\tfrac{\beta^m}{\sqrt{m!}} \ket{n,m}_{ab}.
$
As described previously, the BS transforms the input pair of dual basis FS to
$\ket{n,m}_{ab}\overset{U(\theta)}{\longrightarrow} \ket{n,m}^{(out)}_{ab} = \sum_{p=0}^n f^{(n,m)}_{p}(\theta)\ket{p,n+m-p}_{ab}$.
Inserting this into the above and projecting onto $\ket{m_a, m_b}_{ab}$ for the measurement of $m_a$ photons in mode $a$ and $m_b$ photons in mode $b$ (which produces the delta functions $\delta_{p,m_a}$ and  $\delta_{m,m_a+m_b-n}$), and performing the sum over $p$ and $m$
we obtain
\bea{Prob:odd:cat:state}
\hspace{-0.3in}
&{}& P(m_a,m_b,\theta) = \left(\frac{2}{\mathcal{N}}\right)^2\,e^{-(|\alpha|^2+|\beta|^2)}\, \no
\hspace{-0.3in}
&\times&
\left| 
\sum_{n=0}^\infty 
\frac{\alpha^n\,s_n}{\sqrt{n!}}\frac{\beta^{m_a+m_b-n}}{\sqrt{(m_a+m_b-n)!}} \, f^{(n,m_a+m_b-n)}_{m_a}(\theta)
\right|^2
\overset{\theta=\pi/2}{\underset{m_a=m_b}{\longrightarrow}} 0. \qquad 
\eea
This occurs because the factor of $s_n$ from the odd cat state ensures there are  only odd values of $n$ present, and \tit{each} BSC 
$f^{(n,m_a+m_b-n)}_{m_a} \propto g(m_a,m_b,\theta|n\to odd)\overset{\theta=\pi/2}{\underset{m_a=m_b}{\longrightarrow}} 0$, as discussed above. Note that the amplitudes of the input coherent state for $\ket{\beta}_b$ just \tit{go along for the ride} here. Neither was it crucial what the amplitudes for the input state in mode $a$ were. What \tit{was} critical was the fact the one of the input states to the BS (here the $a$ mode) had \tit{only odd number of FS}, which then plucked out BSCs 
$f^{(n,m=m_a+m_b-n)}_{p}(\theta)\propto g(m_a,m_b,\theta|n)$ with \tit{odd} $n$, each of which goes to zero for a 50:50 BS and for (number resolved) coincident measurements with equal number of photons in each mode, i.e. $m_a=m_b$.
That the CNL for a 50:50 BS with $m_a=m_b$ measurements is a general, universal result is illustrated in 
\Fig{fig:rb:plots:18Aug2021} for the representative case of a CS of amplitude $\beta=3$ ($\bar{n}_b=|\beta|^2$) in mode $b$ mixed with various $a$ mode states containing only odd number of photons, including
(top left) $\ket{3}_a$ (the solid black diagonal line is the CNL),
(top right) $(\ket{1}_a+\ket{3}_a)/\sqrt{2}$, 
(bottom left) photon-added single mode squeezed state:   $a^\dag \ket{SMSS}_a$, and 
(bottom right) the odd cate state $\ket{\alpha_c}_a$.


The main results of this section can be summarized as follows:
for the measurement of the joint probability distribution $P(m',m',\pi/2|n)$ 
with equal number of photons in both output modes of a 50:50 BS 
we have
\begin{itemize}
  \item for $n=2n'+1$ \tit{odd}: \quad $P(m',m',\pi/2|2n'+1)\propto g^2(m,m,\pi/2|2n'+1)\equiv 0$, \tit{regardless} of the input state in mode $b$, pure or mixed. We term this a (diagonal) \tit{central nodal line} (CNL) 
  (see \Eq{gmmtheta:n:odd}),
  \item for $n=2n'$ \tit{even}: \qquad $P(m',m',\pi/2|2n')\propto g^2(m,m,\pi/2|2n') \ne 0$,\; (see \Eq{gmmtheta:n:even}).
\end{itemize}

\section{Pseudo-Nodal Curves (PNC) for a 50:50 and non-50:50 BS}\label{sec:PNC}
As we have seen in the previous section, central nodal lines (CNLs) exists for 
$a$-mode input states containing superpositions of odd numbered FS
(or individual $n$ odd FS)
when detecting $m_a=m_b$ for a 50:50 BS with $\theta = \pi/2$.
These are the CNL in the bottom rows of
\Fig{fig:FSCS:FSTS:plots:n0:1} for $P(m_a,m_b|n=1)$ and
\Fig{fig:FSCS:FSTS:plots:n2:3} for $P(m_a,m_b|n=3)$
for a FS $\ket{n}_a$ mixed with a (left column) CS $\ket{\beta}_b$ and (right column) TS $\rho_b$.

In addition to the CNL, we also observe in  
\Fig{fig:FSCS:FSTS:plots:n0:1}, \Fig{fig:FSCS:FSTS:plots:n2:3} and \Fig{fig:rb:plots:18Aug2021}
what appear to be \tit{pseudo-nodal curves} (PNC) (e.g. the dashed black lines in 
\Fig{fig:rb:plots:18Aug2021}(top left) passing through the blue dots, the later of which have value zero) 
symmetrically placed about the \tit{center line} (diagonal) $m_a=m_b$ 
(e.g. solid black line in \Fig{fig:rb:plots:18Aug2021}(top left)) of the joint output photon number distribution for all $n> 0$, \tit{even or odd}, when in a 50:50 BS configuration. 
We now seek to understand and quantify their origin.
Without loss of generality we will explore the case of an initial FS/CS input 
$\ket{\Psi_{in}}_{ab} = \ket{n}_a\ket{\beta}_b$  for $n=2, 3$,  
\Fig{fig:FSCS:FSTS:plots:n2:3} (left column) 
and \Fig{fig:rb:plots:18Aug2021}(top left) for $n=3$,
and recall that 
$P(m_a,m_b,\theta)\propto \big( g(m_a,m_b,\theta) \big)^2$.

For a dual FS basis input state $\ket{n,m}_{ab}$ the expressions for $g(m_a, m_b, \theta|n)$ for $n=2, 3$ are given by
\bwt
\bsub
\bea{gmambthetan2}
\hspace{-0.5in}
g(m_a, m_b, \theta|2) &=& 
m_a (m_a-1) \big(\cos^2(\theta/2) \big)^2  
- 2 m_a m_b \big(\cos^2(\theta/2) \big) \big(\sin^2(\theta/2) \big)
+m_b (m_b-1) \big(\sin^2(\theta/2) \big)^2, \qquad \\
&\overset{\theta\to\pi/2}{\longrightarrow}& \frac{1}{4}\,\big( m_a (m_a-1) - 2 m_a m_b + m_b (m_b-1) \big),  \label{gmambthetan2:pidiv2}\\
&\overset{\theta\to\pi/3}{\longrightarrow}& 
\frac{1}{16}
\big(9 m_a (m_a-1) - 3 m_a m_b + m_b (m_b-1) \big), \label{gmambthetan2:pidiv3}
\eea
\bea{gmambthetan3}
\hspace{-0.5in}
g(m_a, m_b, \theta|3) &=& 
m_a (m_a-1) (m_a-2)\big(\cos^2(\theta/2) \big)^3  
- 3 m_a (m_a-1)  m_b \big(\cos^2(\theta/2) \big)^2 \big(\sin^2(\theta/2) \big), \no
&+& 3 m_a   m_b (m_b-1)  \big(\cos^2(\theta/2) \big) \big(\sin^2(\theta/2) \big)^2
+ m_b (m_b-1) (m_b-2) \big(\sin^2(\theta/2) \big)^3, \\
&\overset{\theta\to\pi/2}{\longrightarrow}& 
\frac{1}{8}\,\big(
m_a (m_a-1) (m_a-2)
- 3 m_a (m_a-1)  m_b
+ 3 m_a   m_b (m_b-1) 
- m_b (m_b-1) (m_b-2) 
\big), \label{gmambthetan3:pidiv2}\\
&\overset{\theta\to\pi/3}{\longrightarrow}& 
\frac{1}{64}
\big(
   27 m_a (m_a-1) (m_a-2)
- 27 m_a (m_a-1)  m_b
+ 9 m_a   m_b (m_b-1) 
- m_b (m_b-1) (m_b-2)
\big), \qquad\quad \label{gmambthetan3:pidiv3}
\eea
\esub
\ewt
where we have evaluated the expressions for the two angles $\theta=\pi/2$ (50:50 BS) and $\theta=\pi/3$ (non-50:50 BS).

\subsection{PNC for a 50:50 BS}\label{subsec:50:50:BS}
A necessary condition for the existence of solutions of $g(m_a, m_b, \theta|n)=0$  for integer values of $(m_a,m_b)$ is that the above polynomials in $m_a, m_b$ in have \tit{integer coefficients}, which occur when $T=\cos^2(\theta/2)$, and hence $R=\sin^2(\theta/2)$, are \tit{rational numbers}, i.e. a fraction given by the ratio of integers. For the cases chosen  
$\theta=\pi/2$ (50:50 BS) yields $\left(\cos^2(\theta/2), \sin^2(\theta/2)\right)=(1/2,1/2)$, and for  
$\theta=\pi/3$ (non-50:50 BS) $\left(\cos^2(\theta/2), \sin^2(\theta/2)\right)=(3/4, 1/4)$.
Polynomial equations with integer coefficients and integer solutions are known as \tit{Diophantine polynomials}.
Isolated zeros of these equations can be found by a simple brute force search (BFS) over integer values of $(m_a, m_b)$.
Analyltic solutions can be found by various different methods, one of which is by assuming a parametric form of  $(m_a(k),m_b(k))$ in terms of another integer $k\in\mathbb{Z}$. 
For a 50:50 BS ($\theta=\pi/2$) and low values of $n$, we have found that quadratic polynomials in $k$ of the parametric form
$m_a(k)=a_0 + a_1\,k + a_2\,k^2$ and $m_b(k)=b_0 + b_1\,k + b_2\,k^2$ with $\{a_0, a_1, a_2, b_0, b_1, b_2\} \in \mathbb{Z}$ again integers, are sufficient to find analytic solutions for the PNC in the case of $n=2$ and $n=3$ for a 50:50 BS,
i.e. $P\big(m_a(k), m_b(k), \pi/2|n=2,3\big)\equiv 0$, identically.

\begin{figure}[ht]
\begin{center}
\begin{tabular}{ccc}
\hspace{-0.25in}
\includegraphics[width=1.5in,height=1.75in]{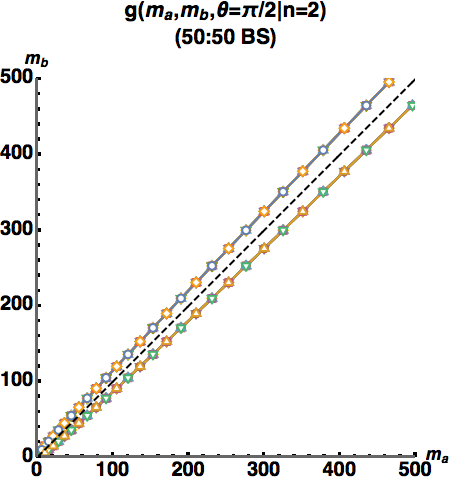} 
& {}\qquad\qquad{} &
\hspace{-0.25in}
\includegraphics[width=1.5in,height=1.75in]{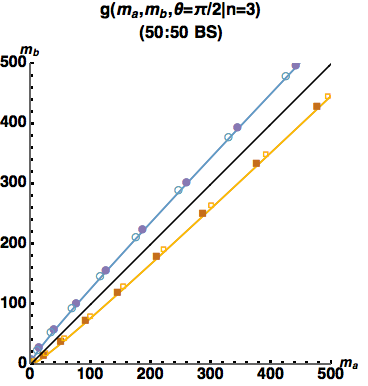}
\end{tabular}
\caption{\label{fig:gmambpidiv2n2n3_BruteForceOnlyn2_mambofk} 
Plots of zeros of 
(left)   $g(m_a,m_b,\theta=\pi/2|n=2)$, 
(right) $g(m_a,m_b,\theta=\pi/2|n=3)$ 
for a 50:50 BS with $\theta=\pi/2$.
Plot markers: 
(solid)  ``upper and lower branch" solutions analytically found by \ttt{Reduce}  \cite{Reduce:comment}
in \tit{Mathematica},
(open) ``upper and lower  branch" solutions found by a BFS over approximately 
(left) $11.4\times 10^{6}$,
(right) $9.5\times 10^{8}$
$6$-tuples integer coefficients 
$(a_0, a_1, a_2, b_0, b_1, b_2)$
of two quadratic polynomials (in $k$, integer)
for $(m_a = a_0 + a_1\,k + a_2\,k^2, m_b = b_0 + b_1\,k + b_2\,k^2)$ which identically solve $g(m_a,m_b,\theta=\pi/2|n)=0$ analytically
(see Table~\ref{tbl:n2:theta:pidiv2} and Table~\ref{tbl:n3:theta:pidiv2} in \App{tables:ma:mb:of:k}).
Note: The (left) dashed black line is the diagonal drawn for symmetry for $n=2$, 
while
(right) the solid black line is the CNL for $n=3$.
}
\end{center}
\end{figure}

In \Fig{fig:gmambpidiv2n2n3_BruteForceOnlyn2_mambofk} we plot the case of a 50:50 BS($\theta=\pi/2$) for
 (left) $g(m_a, m_b, \theta=\pi/2|n=2)$, and (right) $g(m_a, m_b, \theta=\pi/2|n=3)$ with the 
plot markers indicating the zeros arising from polynomials  $\big(m_a(k),m_b(k)\big)$  (for $m_a\ne m_b$), 
which lie on the solid curves (the PNC).
An example of such a parametric solutions for $n=2$ is $\big(m_a(k),m_b(k)\big) = (2 k^2 -k,\, 2k^2 -3k +1)$, 
while for 
$n=3$ an example is $\big(m_a(k),m_b(k)\big) = (6 k^2 +7 k +2,\, 6 k^2 +13 k +7)$
(see Table~\ref{tbl:n2:theta:pidiv2} and Table~\ref{tbl:n3:theta:pidiv2}, respectively in \App{tables:ma:mb:of:k}).
%
%
 The black dashed central (symmetry) line represents a CNL solution, ($m_a=m_b$) only for $n=3$. For $n=2$ we have drawn it in as a diagonal in order to see the symmetric behavior of the plotted points. In both case $n=2,3$, the colored solid lines in \Fig{fig:gmambpidiv2n2n3_BruteForceOnlyn2_mambofk} are the PNC on which  the  (isolated, non-contiguous) zeros (indicated by the plot markers) of  $g(m_a,m_b,\theta=\pi/2|n=2,3)$
 lie that were generated from the quadratic polynomial solutions $(m_a(k),m_b(k))$. 

 For the case of $n=3$ we have overlaid 
 the CNL (solid black diagonal line of complete destructive interference $g(m',m',\pi/2|n=3)=0$), 
 and the PNC (dashed black non-diagonal curve) of 
 \Fig{fig:gmambpidiv2n2n3_BruteForceOnlyn2_mambofk}(right)
 onto the contour plot of 
 \Fig{fig:rb:plots:18Aug2021}(top left).
 The blue dots in the latter plot are the points $(m_a, m_b\ne m_a)$ of complete destructive interference where 
 $g(m_a,m_b,\pi/2|n=3)=0$ through which the PNC (dashed black curves) pass. Along the PNC, in between the zero values (blue dots), $g(m_a,m_b,\pi/2|n=3)$ obtains nearly (but not) complete destructive interference (``fringes"), i.e. extremal values. This later statement is most clearly illustrated by explicitly examining the case of $n=2$ for the input state $\ket{2}_a\ket{\beta}_b$, 
 \Fig{fig:FSCS:FSTS:plots:n2:3}(top left), and \Fig{fig:n2:contour:wPNCs} below.
\begin{figure}[ht]
\begin{center}
\includegraphics[width=3.0in,height=2.25in]{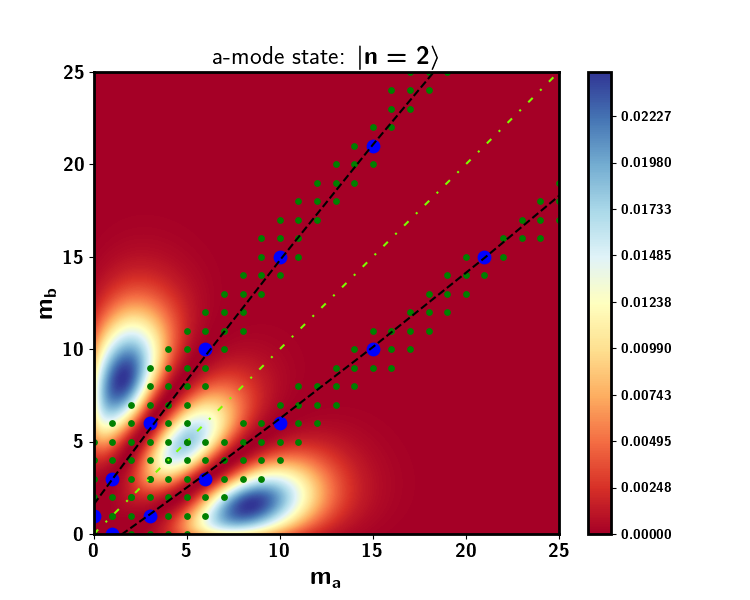}
\caption{\label{fig:n2:contour:wPNCs} 
Contour plot of the
output joint probability $P(m_a,m_b,\theta=\pi/2)$ in 
 \Fig{fig:FSCS:FSTS:plots:n2:3}(top left)
for input state  $\ket{2}_a\ket{\beta}_b$
 with PNC (non-diagonal black dashed curves).
The blue dots are points for which $g(m_a,m_b,\pi/2|n=3)\equiv 0$ through which the PNC pass.
The green dots are points for which $dg(m_a,m_b,\pi/2|n=3)/dm_a = 0$ (extremal values of near, but not complete, destructive interference). The  yellow, dot-dashed line is drawn in to highlight the  diagonal symmetry.
}
\end{center}
\end{figure}

For $n=2$ we have 
$g(m_a,m_b,\pi/2|n=2) =1/4\,\big( m_a (m_a-1) - 2 m_a m_b + m_b (m_b-1) \big)$, \Eq{gmambthetan2:pidiv2}.
Temporarily treating $m_a$ and $m_b\equiv m_b(m_a)$ as continuous variables, let us formally solve
$dg(m_a,m_b,\pi/2|n=2)/dm_a=0$ for $\tfrac{dm_b}{dm_a} = \tfrac{2(m_b-m_a)+1}{2(m_b-m_a)-1}$.
Solving this latter ordinary differential equation for $m_b(m_a)$ yields
$m^{(\pm)}_b = m_a + \tfrac{1\pm\sqrt{1+ 8 m_a + k}}{2}$ where $k\in\mathbb{Z}$ is an arbitrary integration constant, such that for $m_a\in\mathbb{Z}_{\ge0}$ we accept only integer solutions of $m_b\in\mathbb{Z}_{\ge0}$. 
(An obvious example solution is $(m_a=1,k=0)$ leading to solutions $(m_a=1,m^{(+)}_b=3)$ and  $(m_a=1,m^{(-)}_b=0)$).
These are the green dots in \Fig{fig:n2:contour:wPNCs}, 
($(m_a, m^{(+)}_b)$, $(m_a, m^{(-)}_b)$ upper and lower branch, respectively),
the center of which where the PNC lie, 
as well as the exact zeros (blue dots) of $g(m_a,m_b,\pi/2|n=2)$.
While  $dg(m_a,m_b,\pi/2|n=2)/dm_a$ is not a proper derivative (since $m_a, m_b$ are discrete), 
the points along which it takes zero values (green dots) are 
extremal points of near, but not complete, destructive interference (``fringes"), i.e. ``valleys" of the output joint probability distribution. 
The ``valley floor" is the PNC (dashed black lines) of minimal values, 
along which are scattered exact zeros (blue dots) of $g(m_a,m_b,\pi/2|n=2)$
representing non-diagonal points of complete destructive interference.
Thus, we can say for PNC of general $n$, that 
$\{ (m_a, m_b) \,|\, g(m_a,m_b,\pi/2|n) = 0\} \subset \{\, (m_a, m_b)\,|\, dg(m_a,m_b,\pi/2|n)/dm_a = 0\}$
(i.e. blue dots $\subset$ green dots). 
These PNCs furcate (multi-divide) the output joint probability distribution into peaks and carved out valleys, as evident
in \Fig{fig:FSCS:FSTS:plots:n0:1} and \Fig{fig:FSCS:FSTS:plots:n2:3}.

\subsection{PNC for a non-50:50 BS}\label{subsec:non:50:50:BS}
\begin{figure}[ht]
\begin{center}
\begin{tabular}{ccc}
\hspace{-0.25in}
\includegraphics[width=1.75in,height=1.75in]{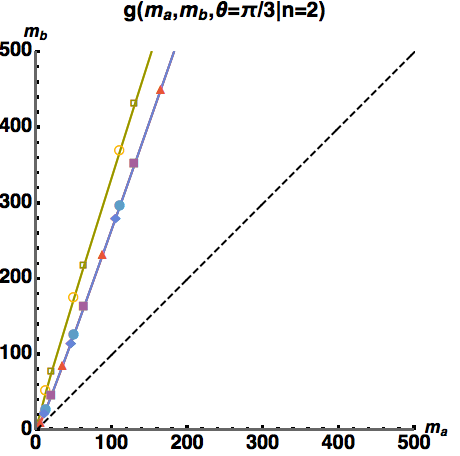} 
&{}&
\hspace{-0.45in}
\includegraphics[width=2.5in,height=1.75in]{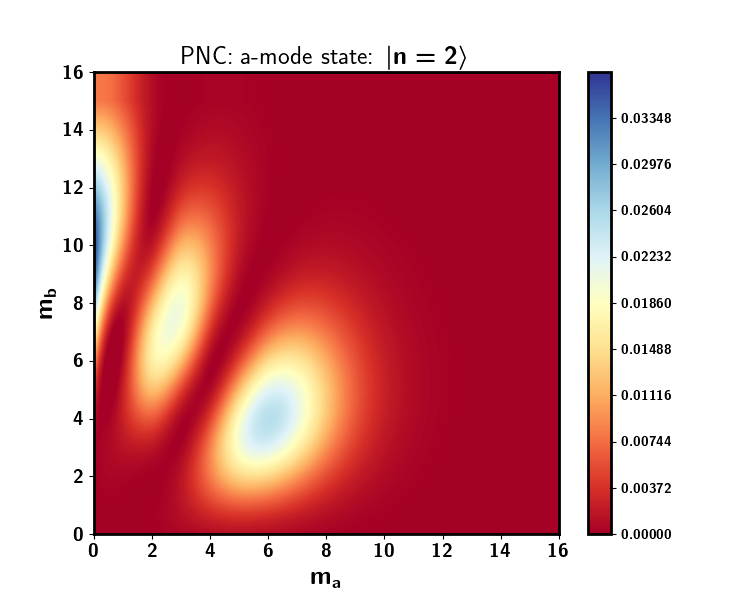} 
\end{tabular}
\caption{\label{fig:gmambpidiv3n2_analytic_solns_mambofk} 
Plots of zeros of $g(m_a,m_b,\theta=\pi/3|n=2)$ 
for a \tit{non}-50:50 BS with $\theta=\pi/3$.
Plot markers: 
(solid)  ``upper branch" solutions analytically found by \ttt{Reduce} in \tit{Mathematica},
(open) ``lower branch" solutions found by BFS over approximately $9.6\times 10^{8}$ $6$-tuples integer coefficients 
$(a_0, a_1, a_2, b_0, b_1, b_2)$
of two quadratic polynomials (in $k$, integer)
for $(m_a = a_0 + a_1\,k + a_2\,k^2, m_b = b_0 + b_1\,k + b_2\,k^2)$ which identically solve $g(m_a,m_b,\theta=\pi/3|n=2)=0$ analytically.
See Table~\ref{tbl:n2:theta:pidiv3} of \App{tables:ma:mb:of:k}.
(right) Contour plot blowup of lower $(m_a, m_b)$ pairs region for a coherent $b$-mode input state.
}
\end{center}
\end{figure}

The question naturally arises whether such polynomial solutions $(m_a(k),m_b(k))$ also exist for non-50:50 BS angles $\theta\ne \pi/2$ again with $\big(T=\cos^2(\theta/2), R=\sin^2(\theta/2)\big)$ rational numbers. The primary difference in the form of the Diophantine equations for arbitrary BS angles is now the integer coefficients are not \tit{alternating symmetric} 
(i.e. of the binomial form $\binom{n}{q}\,(-1)^q$) as in the 50:50 BS case).
In \Fig{fig:gmambpidiv3n2_analytic_solns_mambofk}(left) we plot the PNC for 
$g(m_a, m_b, \theta=\pi/3|n=2)$ for a non-50:50 BS angle $\theta=\pi/3$, 
again with the colored solid lines  indicating the PNC on which 
\begin{figure}[ht]
\begin{center}
\includegraphics[width=3.0in,height=2.25in]{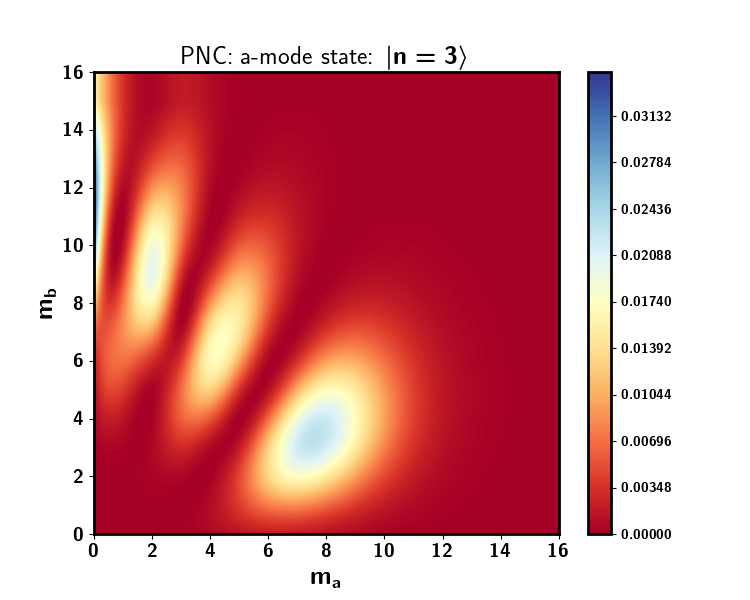}
\caption{\label{fig:gmambpidiv3n3_analytic_solns_mambofk} 
Same as  \Fig{fig:gmambpidiv3n2_analytic_solns_mambofk}(right), except for $n=3$;\; 
$g(m_a,m_b,\theta=\pi/3|n=3)$. 
}
\end{center}
\end{figure}
 the  (isolated, non-contiguous) zeros (indicated by the plot markers) of  $g(m_a,m_b,\theta=\pi/3|n=2)$ lie which
 are generated from the quadratic polynomial solutions $(m_a(k),m_b(k))$. 
 An example of such a parametric solutions for $n=2$ is $\big(m_a(k),m_b(k)\big) = (12 k^2 +k,\, 36k^2 -9 k)$ 
(see Table~\ref{tbl:n2:theta:pidiv3}  in \App{tables:ma:mb:of:k}).
 The (non-CNL) black dashed line $m_a=m_b$ is drawn to simply draw attention to the rotation of the PNC from 
 this diagonal symmetry line from the previous 50:50 BS solution. 
  \Fig{fig:gmambpidiv3n2_analytic_solns_mambofk}(right) is a contour plot blowup of lower $(m_a, m_b)$ pairs region.
 
Finding analytic solutions for  $g(m_a, m_b, \theta=\pi/3|n=3)$ for the same non-50:50 BS angle of $\theta=\pi/3$ 
\Fig{fig:gmambpidiv3n3_analytic_solns_mambofk},
has proven to be problematic.
Isolated zeros are easily found by a BFS double looping over $m_a, m_b\in\big\{1,\ldots,10^4\big\}$ producing
$(m_a, m_b) \in \{(1,0), (1,1), (1,11), (2,0), (3,1), (11,55), (70,162)\}$.
The candidate solutions 
$(m_a, m_b) \in \{(1,0), (1,1), (2,0)\}$ are trivially non-physical due to the delta function constraint $\delta_{m_a+m_b,n+m}$ since $n+m\ge 3$ for the FS input state $\ket{3}_a$.
A BFS over quadratic polynomials of the form 
$m_a(k)=a_0 + a_1\,k + a_2\,k^2$ and $m_b(k)=b_0 + b_1\,k + b_2\,k^2$ with $\{a_0, a_1, a_2, b_0, b_1, b_2\} \in \mathbb{Z}$
with each coefficient $a_k, b_k$ in the search range $\{-10, 50\}$ (i.e. $51.5 \times 10^9$ 6-tuples) failed to produce any analytic solutions. 
Similarly, another BFS over cubic polynomials for the form 
$m_a(k)=a_0 + a_1\,k + a_2\,k^2 + a_3\,k^3$ and 
$m_b(k)=b_0 + b_1\,k + b_2\,k^2+ b_3\,k^3$ with $\{a_0, a_1, a_2, a_3, b_0, b_1, b_2, b_3\} \in \mathbb{Z}$
with each coefficient $a_k, b_k$  in the search range $\{-10, 10\}$ (i.e. $37.8 \times 10^9$ 8-tuples) also failed to produce any analytics solutions. These numerical searches for exact polynomial solutions $\big(m_a(k), m_a(k)\big)$ of parametric form in 
$k\in\mathbb{Z}$ clearly do not exclude the possibility for other forms of analytic solutions
%
(e.g. solutions of cubic equations, if one naively solves 
$g(m_a,m_b,\pi/3|n=3)=0$ for $m_b(m_a)$ as a function of $m_a$, requiring $m_b\in\mathbb{Z}_{\ge 0}$  \cite{Reduce:comment}).

As a conjectured explanation for the failure to find many zeros for $g(m_a, m_b,\pi/3|n=3)$ 
it should be noted that as we rotate the BS angle from $\theta=\pi/2$ (50:50) to either extremes $\theta=0$ ($T=1$), 
or $\theta=\pi$ ($R=1$) the polynomial
 $g(m_a, m_b,\theta|n)$ for general $\theta$  and $n$ degenerates to either $(m_a)_n$ or $\pm(m_b)_n$ respectively, since 
 $\big(\cos^2(\theta/2), \sin^2(\theta/2)\big)\to \{(1,0), (0,1)\}$, respectively, where no PNC exist.
 That is, as we rotate $\theta$ any PNC that lie symmetrically placed about the CNL ($n$ odd) or diagonal $m_a=m_b$ ($n$ even) must eventually vanish. We conjecture that this vanishing of the PNC for non-50:50 BS angles happens more severely in $g(m_a, m_b,\theta|n)$ for $n\ge 3$.

\section{Prospects for experimental realization}\label{sec:Experimental}
The potential for experimental realization of the CNL and PNC discussed in this work rests on the ability to perform \tit{photon number-resolving} detection. Most detectors such as avalanche photodiode detectors (APDs) are ``bucket" detectors meaning that they either record no signal (in the absence of spurious dark counts - which are always present, but quantifiable) 
or a single ``click" from the collection of photons that enter the detector in the sampling window. Such \tit{click/no-click} detectors can be  multiplexed to achieve a quasi-photon number resolution  \cite{Gerrits:2016:v2}. However, in those cases the fidelity of the measured state is always degraded compared to true/intrinsic photon-number-resolving detection. In many applications, however, multiplexed click detectors are sufficient to achieve high-fidelity state characterization  \cite{Gerrits:2016:v2}. 

Since 1998, great progress has been made in the development of true number-resolving detectors, specifically
superconducting \tit{transition edge sensors} (TES). In contrast to simple ‘click/no-click’ detectors, the TES output contains information about the number of photons absorbed. TES are highly-sensitive microcalorimeters that are used as microbolometers to detect radiation from sub-mm wavelengths to gamma-rays. From the review article by Gerrits \tit{et al.}  \cite{Gerrits:2016:v2}, the optical TES is a superconducting sensor measuring the amount of heat absorbed from an optical photon with energy on the order of 1 eV. When an optical photon is absorbed by the sensor, the associated photon energy is transformed into a measurable temperature change of the sensor. Typical TES operate at temperatures below 1K. Transition edge sensors typically resolve photons over a range of 1 to 10 or 20, depending on the wavelength, heat capacity of the device, and steepness of the superconducting resistance transition.

Germane to our considerations Gerrits \tit{et al.} (2016) \cite{Gerrits:2016:v2} have measured a coherent state with mean photon number $|\beta|^2\approx 5$
 using an optimum filter analysis  to determine the pulse-height histogram. 
 Well defined non-overlapping peaks for $0-4$ photons
 are clearly discernible (see Fig.2.8(c) of  \cite{Gerrits:2016:v2}). 
 Recently, Schmidt \tit{et al.} (2018)  \cite{Schmidt:2018}
using quantum dot (QD)-based photon sources emitting at 932nm (1.33eV), have reported that adjacent photon number states up to 25 photons can be discriminated with the TES detectors 
with efficiencies exceeding $87\%$
using principal component (PC) analysis 
 (see Fig.(3) of  \cite{Schmidt:2018}). 
 In the optical regime exemplary analog signal output of $0-6$ photons can be clearly distinguished by pulse heights at wavelength of 653 nm (1.9 eV) (see Fig.(1, bottom, middle) of  \cite{Schmidt:2018}). 
 In these experiments 
 each photon state in the photon number distributions obtained from the pulse area and the PC analyses is fitted with a Gaussian function. The corresponding full width at half maximum (FWHM) is interpreted as the energy resolution  $\Delta E_{FWHM}$ of the TES detectors.
 
 Thus, the prospect to observe the predicted CNL and PNC by histograming the photon number-resolved  output of the BS 
 in the range $(m_a,m_b)\in\{0,1,2,3,4,5\}$ seems very promising. This, of course, assumes that one can prepare $a$-mode input states with only odd numbers of photons (odd parity state), such as odd numbered FS, photon-added or photon-subtracted single-mode squeezed states, or e.g. an odd (coherent state-based) cat state. The most promising of these prospects that appears most readily achievable with today's current technology is an input state consisting of a single photon FS  $\ket{1}_a$  in the $a$-mode (obtained by heralding on either the signal or idler component of a biphoton, two-mode squeezed state),
mixed with either a weak CS, or possibly a weak TS, 
(see second row of \Fig{fig:FSCS:FSTS:plots:n0:1})
with a mid-range mean photon number around $\bar{n}\approx 3$.

Realistic experiments will always involve the use of imperfect detectors.
The inclusion of loss is easily performed by averaging the joint photon number distribution $P(m_a,m_b,\theta)$ over a Bernoulli distribution with a Beer's law loss parameter (detector efficiency, probability) given by $0\le\eta=e^{-2\gamma\,L}\le 1$ as discussed in LBK \cite{Lai:1991}, and more recently in Laiho \tit{et al.} \cite{Laiho:2019}. The basic idea is that if $m_a$ photons are detected in the $a$-mode output of the BS, it could have resulted from a total of 
$M_a\ge m_a$ photons, all but which $m_a$ were lost, and hence not registered by the detector. The same is true for the detection of $m_b$ photons in the $b$-mode output of the BS.

Let us define the $a$-mode projection operator
\be{loss:opr}
\Pi_a = \sum_{M_a\ge m_a}^\infty \binom{M_a}{m_a}\,\eta^{m_a} \, (1-\eta)^{(M_a-m_a)}\,\ket{M_a}\bra{M_a},
\ee
and an analogous expression for $\Pi_b$. The Bernoulli  factor in the summation is the probability for $m_a$ $a$-mode photons to traverse the BS with 
detection probability $\eta$ per photon, 
with the remaining $(M_a-m_a)$ suffering loss (non-detecton) with probability $(1-\eta)$. 
These latter $(M_a-m_a)$ ``lost'' photons can be considered as having been ``reflected" into a scattering mode 
with probability $R=(1-\eta)$
which is then not observed (e.g. via an additional ``virtual BS" with transmissivity $T=\eta$, and $R=(1-\eta)$,
see Loudon \cite{Loudon:1997}, and Appendix C of \cite{Alsing_Hach:2017a}).
Note that as the detector efficiency goes to unity $\eta\to1$, implying no loss, 
the factor $(1-\eta)^{(M_a-m_a)}$ is non-zero only when $M_a=m_a$ and the projection operator collapses to 
$\Pi_a\overset{\eta\to 1}{\longrightarrow} \ket{m_a}_a\bra{m_a}$.

\begin{figure*}[ht]
\begin{center}
\begin{tabular}{ccc}
\hspace{-0.5in}
\includegraphics[width=2.5in,height=2.25in]{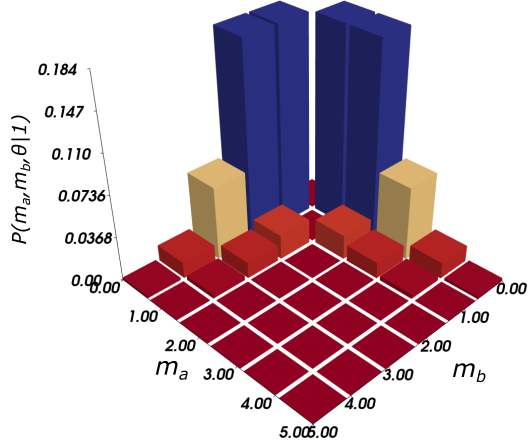} & {\hspace{0.5in}} &
\includegraphics[width=2.5in,height=2.25in]{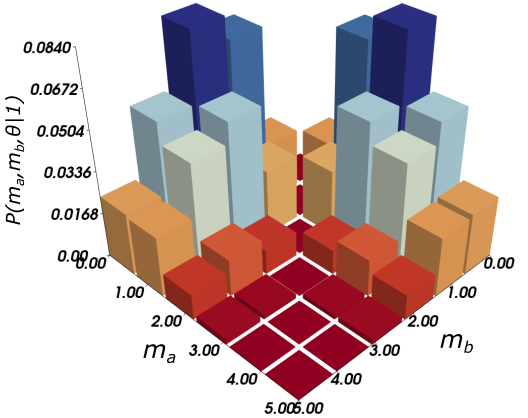} \\
\includegraphics[width=2.25in,height=0.65in]{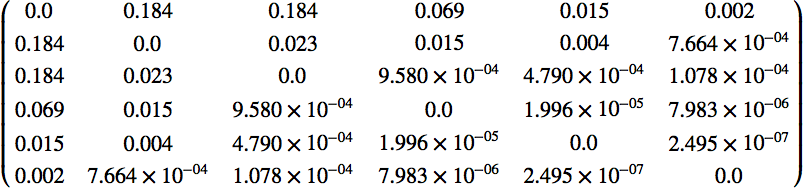}  & {\hspace{0.5in}} &
\includegraphics[width=2.25in,height=0.65in]{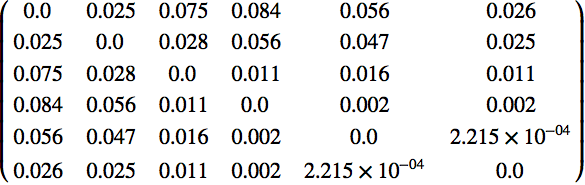}
\end{tabular}
\caption{\label{fig:FSn1:CS:beta:1:root3} 
Output joint probability distribution $P(m_a,m_b,\pi/2|n=1)$ for a FS/CS input 
$\ket{\Psi_{in}}_{ab} = \ket{1}_a\ket{\beta}_b$
for a 50:50 BS with mean number of photons in the $b$ mode $\bar{n}_b = |\beta|^2$:
(top row, left) $\bar{n}_b =1$, 
(top row, right) $\bar{n}_b =3$.
The bottom row indicates the probability matrix $P(m_a,m_b,\pi/2|n=1)$ for the plot in the corresponding column,
with rows $m_a$ and columns $m_b$ labeled by the indices $(m_a,m_b)\in\{0,1,2,3,4,5\}$.
}
\end{center}
\end{figure*}

The output joint probability distribution $\tilde{P}(m_a,m_b,\theta;\eta_a, \eta_b)$ of the BS, 
with  efficiencies $\eta_a, \eta_b$ in modes $a$ and $b$ respectively, is now given by
\bea{Pmambtheta:with:loss}
\hspace{-0.15in}
\tilde{P}(m_a,m_b,\theta;\eta_a, \eta_b) &=& \Tr_{ab}[\Pi_a\otimes\Pi_b\,\rho^{(out)}_{ab}], \no
= 
 \sum_{M_a\ge m_a}^\infty 
 \sum_{M_b\ge m_b}^\infty
 &{}&
 \eta_a^{m_a}\, (1-\eta_a)^{(M_a-m_a)} \, \times \no
 &{}&
  \eta_b^{m_b}\,(1-\eta_b)^{(M_b-m_b)} \,
 P(M_a,M_b,\theta). \qquad
\eea
%
In \Fig{fig:FSn1:CS:beta:1:root3} we plot the 
output joint probability distribution $P(m_a,m_b,\pi/2|n=1)$ for a 
FS/CS input $\ket{\Psi_{in}}_{ab} = \ket{1}_a\ket{\beta}_b$
for an ideal lossless  50:50 BS with mean number of photons $\bar{n}_b = |\beta|^2$ in the $b$ mode.
The top left plot is for  $\bar{n}_b =1$, while the top right plot is for  $\bar{n}_b =3$.
The probability distribution matrix is plotted directly below in the second row for the corresponding plot immediately above, with rows $m_a$ and columns $m_b$ labeled by the indices $(m_a,m_b)\in\{0,1,2,3,4,5\}$.
Due to the small average photon numbers used in the $a$ and $b$-mode input states, dictated by the 
discriminating capability of current (e.g. TES) photon-number resolving detectors,
by the time we get to $m_a, m_b = 3$ one must be able to distinguish 
a non-zero probability for joint photon detection from zero probability to roughly $1\%$.
While challenging, such measurements are nonetheless within the realm of possibility with today's current technology.
\begin{figure*}[ht]
\begin{center}
\begin{tabular}{ccc}
\hspace{-0.5in}
%
\includegraphics[width=2.5in,height=2.25in]{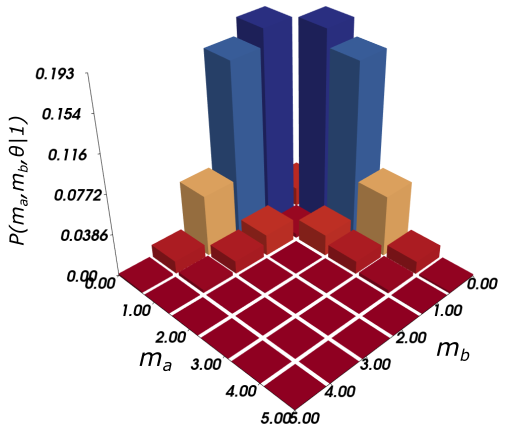} & {\hspace{0.5in}} &
\includegraphics[width=2.5in,height=2.25in]{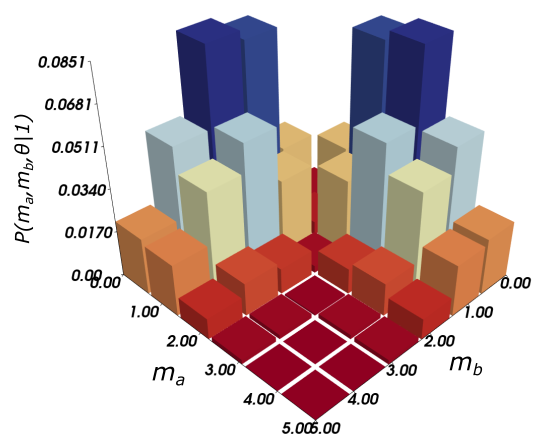} \\
\includegraphics[width=2.25in,height=0.65in]{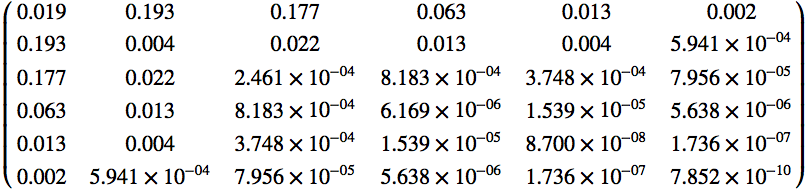}  & {\hspace{0.5in}} &
\includegraphics[width=2.25in,height=0.65in]{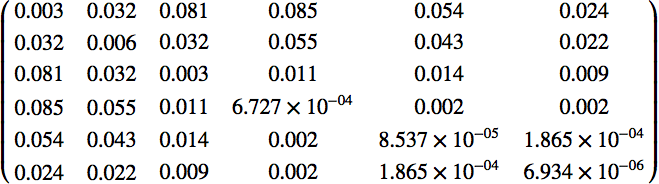}
\end{tabular}
\caption{\label{fig:FSn1:CS:beta:1:root3:loss:eta:0p95} 
Same as \Fig{fig:FSn1:CS:beta:1:root3} but with the inclusion of loss: $\eta_a=\eta_b=\eta=0.95$ ($5\%$ loss) in \Eq{Pmambtheta:with:loss}.
}
\end{center}
\end{figure*}

In \Fig{fig:FSn1:CS:beta:1:root3:loss:eta:0p95} we show the same plot as 
\Fig{fig:FSn1:CS:beta:1:root3}, now with the inclusion of loss, with
$\eta_a=\eta_b=\eta=0.95$ ($5\%$ loss) in \Eq{Pmambtheta:with:loss}.
The diagonal CNL is no longer exactly zero due to the presence of a small amount of loss, but there is still a discernible near central nodal line that furcates the output joint probability distribution. 
It should be noted that losses in quantum integrated waveguide-coupler BS can be as low as $1\%$.

\begin{figure*}[ht]
\begin{center}
\begin{tabular}{ccc}
\includegraphics[width=2.5in,height=2.25in]{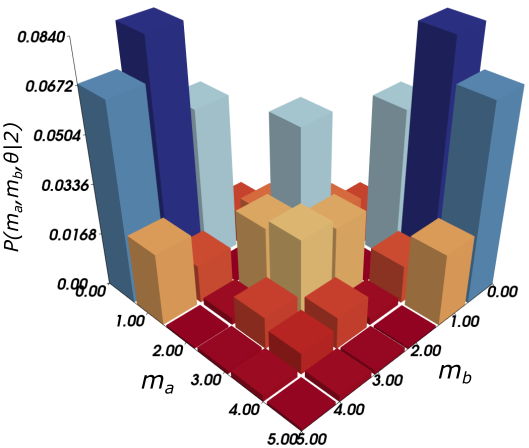} & {\hspace{0.5in}} &
\includegraphics[width=2.5in,height=2.25in]{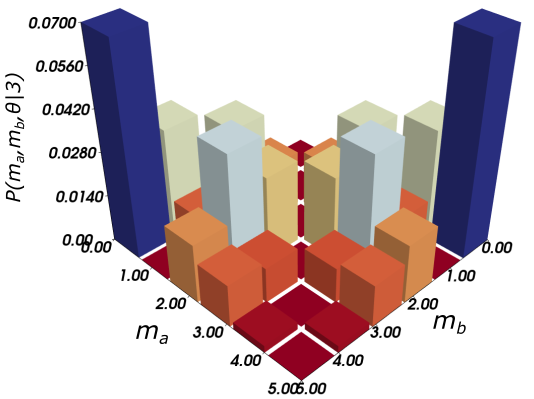} \\
\includegraphics[width=2.5in,height=2.25in]{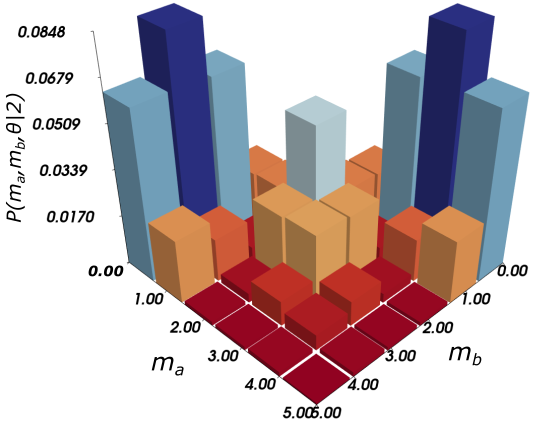}  & {\hspace{0.5in}} &
\includegraphics[width=2.5in,height=2.25in]{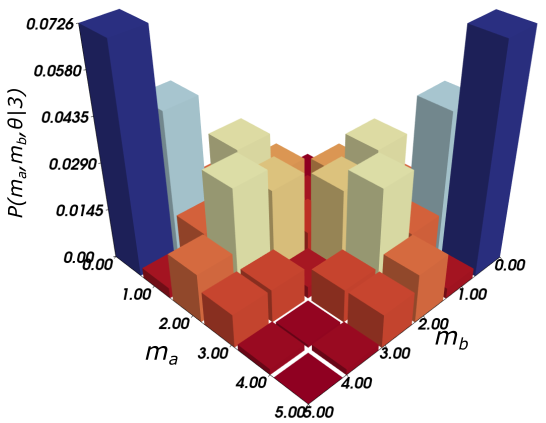}
\end{tabular}
\caption{\label{fig:FSn2n3:CS:beta:root3:loss:eta:1p0:0.05} 
Output joint probability distribution $P(m_a,m_b,\pi/2|n)$  (left column: $n=2$, right column: $n=3$)
with inclusion of loss (detector efficiencies $\eta_a=\eta_b=\eta$ in \Eq{Pmambtheta:with:loss})
for a FS/CS input $\ket{\Psi_{in}}_{ab} = \ket{n}_a\ket{\beta=\sqrt{3}\,}_b$
for a 50:50 BS with mean number of photons in the $b$-mode $\bar{n}_b = |\beta|^2=3$:
(top row): $\eta=1.0$, lossless;
(bottom row): $\eta=0.95$, $5\%$ loss.
}
\end{center}
\end{figure*}

To see the effect of loss on the non-diagonal PNC, we plot in \Fig{fig:FSn2n3:CS:beta:root3:loss:eta:1p0:0.05}
the output joint probability distribution $P(m_a,m_b,\pi/2|n)$ 
for $n=2$ (left column) and $n=3$ (right column)
with inclusion of loss: 
$\eta_a=\eta_b=\eta= \{1.0, 0.95 \}\leftrightarrow \{0\%, 5\%\}$ loss, (top row, bottom row, respectively)
for a FS/CS input $\ket{\Psi_{in}}_{ab} = \ket{n=2,3}_a\ket{\beta=\sqrt{3}\,}_b$
for a 50:50 BS with mean number of photons in the $b$-mode $\bar{n}_b = |\beta|^2=3$.
While the degradation of the CNL for $n=3$ is clearly discernible (column 2) with the inclusion of a small amount of loss,
 one can also observe the increase in the ``valley floor" for the PNC, for both $n=2$ and $n=3$, 
 illustrated in the bottom row with  $5\%$ loss ($\eta = 0.95$), over that of the ideal BS (no loss, $\eta=1.0$) 
 in the top row.
These results indicate that experimental observation of the CNL and PNC in the presence of a small amount of noise  is within the realm of possibility, modulo the difficulty/complexity of creating an $a$-mode input  FS (especially for $n>1$).

The difficulty in observing the  PNC  for the  $\ket{2}_a\ket{\beta}_b$ FS/CS  
in the left column of \Fig{fig:FSn2n3:CS:beta:root3:loss:eta:1p0:0.05} is, of course, the preparation of the $n=2$ $a$-mode FS. One possibility is to first use another canonical HOM setup with a single photon in each of the input ports ($a'$ and $b'$) of a 50:50 BS (generated from two multiplexed spontaneous parametric down conversion (SPDC) sources, each which heralds on its own idler), in order to create two separate signal photons$\ket{1,1}_{a'b'}$. When these two photons enter a 50:50 BS, the output amplitude for the state $\ket{1,1}_{a',b'}$ will suffer complete destructive interference, leading to the output state
$\ket{1,1}_{a'b'}^{(out)} = \big(\ket{2,0}_{a'b'}+\ket{0,2}_{a'b'}\big)/\sqrt{2}$  from \Eq{HOM:calc} . 
If one then heralds on a null measurement \cite{Dakna:1998,Nunn:2021} in the $a'$-mode output, then $2$ photons must  emerge in the $b'$-mode output (and visa versa), which can subsequently be used as the $a$-mode $n=2$ FS input state $\ket{2}_a\ket{\beta}_b$ for the left column of \Fig{fig:FSn2n3:CS:beta:root3:loss:eta:1p0:0.05}.
This null measurement occurs with probability one half, if the time delay between the two $a',b'$ input photons arriving at the first BS is zero. 

The advantage of the above measurement scheme is that is does not require the use of number-resolving photon detection. However, if one does have access to number-resolving photon detection (e.g. a TES discussed above), then a straightforward scheme to generate a $2$ or $3$ photon FS is to simply herald one arm of a (biphoton) two-mode squeezed state (TMSS) $\ket{\xi}_{a'b'}~=~\sum_{n=0}^\infty \sqrt{p^{(\xi)}_n}\,\ket{n,n}_{a',b'}$, where 
$p^{(\xi)}_n =\tanh^{2 n}(r)/ \cosh^{2}(r)$
with squeezing parameter $r>0$. The ``amount of squeezing" (in dB) is given by 
$10\log_{10}\left[(\Delta X)^2/(1/4)\right] = 10\log_{10}\left[e^{-2\,r}\right]$ \cite{Walls_Milburn:1994,Gerry_Knight:2004,Agarwal:2013} where $(\Delta X)^2$ is the variance of the quadrature that undergoes squeezing, and $1/4$ (in these units) is the variance of the vacuum state. Typical values of $r$ for strong squeezed sources are in the range of $r\approx1.4-1.6$ leading to $-12.2$dB to $-13.9$dB of squeezing (for comparison, $-3$dB of squeezing corresponds to $r=0.35$).

We are now interested in the probability to detect the $a'$-mode FS state $\ket{n}_{a'}$ from the 
TMSS source $\ket{\xi}_{a'b'}$ with a number-resolving photon detector with efficiency (detection probability) $\eta$.
Following  Motes \tit{et al.} \cite{Motes:2013} and Nunn \tit{et al.} \cite{Nunn:2021} the conditional probability 
$P_D(t|n')$ to detect $t$ photons given that $n'\ge t$ were actually present is given by
$P_D(t|n') = \binom{n'}{t}\,\eta^{t}\,(1-\eta)^{n'-t}$, which is again the same Bernoulli factor arising from the ``fictitious BS" that was used in \Eq{loss:opr} to model loss. Therefore, the total probability of detecting $t$ photons in the heralding arm of a single TMSS source is given \cite{Motes:2013} by
\bea{Motes:7}
P^{SPDC}_D(t) &=& \sum_{n'=t}^\infty \,P_D(t|n')\,p^{(\xi)}_{n'}, \no
&=&  \sum_{n'=t}^\infty \, \binom{n'}{t}\,\eta^{t}\,(1-\eta)^{n'-t}\,p^{(\xi)}_{n'},
\eea
which is the single mode version of \Eq{Pmambtheta:with:loss}.
For example, $P^{SPDC}_D(2)$ is the probability to detect $2$ photons in the heralding arm of the TMSS source when $n'$ are present, but $(n'-2)$ have been lost (unregistered by the detector), then summed over all possible values of $n'\ge 2$.

\begin{figure}[h]
\includegraphics[width=2.55in,height=2.25in]{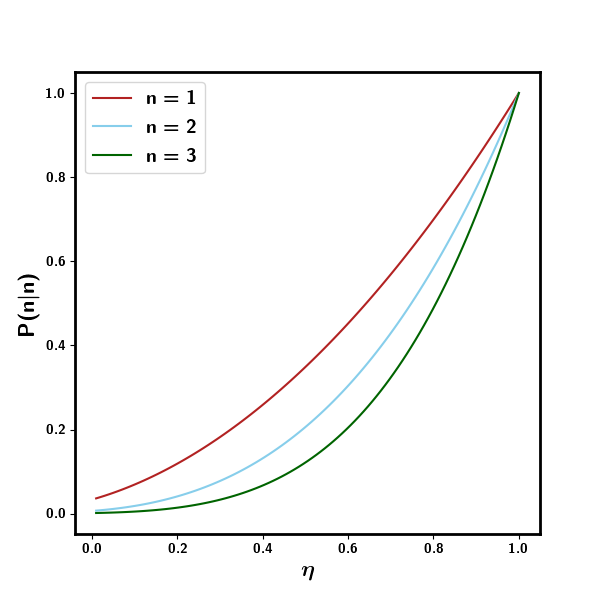} 
\caption{Conditional probability 
$P(n|n)\equiv P_{\ket{n}_{a'}}(n|t=n)$ as a function of detector efficiency $\eta$ for $n~\in~\{1,2,3\}$
that given $n$ photons were detected in the heralding arm (mode $a'$) of the TMSS source, 
it arose from the state $\ket{n}_{a'}$.
}\label{fig:P:n:n}
\end{figure}

Ultimately, we are interested in the inverse conditional probability 
$P_{\ket{n}_{a'}}(n'|t)$
that given $t$ photons were detected in the heralding arm (mode $a'$), these detections actually arose from the input state $\ket{n'}_{a'}$, implying that the output state in the heralded arm is $\ket{n'}_{b'}$.
From Bayes rule we have 
\be{Bayes:rules}
\hspace{-0.155in}
P_{\ket{n'}_{a'}}(n'|t) = \frac{P_D(t|n')\,p^{(\xi)}_{n'}}{P_D(t)} 
= \frac{P_D(t|n')\,p^{(\xi)}_{n'}}{\sum_{n^{''}=t}^\infty \,P_D(t|n^{''})\,p^{(\xi)}_{n^{''}}},
\ee
where the numerator (denominator) in \Eq{Bayes:rules} is the summand (entire sum) in \Eq{Motes:7}.
In \Fig{fig:P:n:n} we plot $P(n|n)\equiv P_{\ket{n}_{a'}}(n|t=n)$ 
as a function of detector efficiency $\eta$
for $n\in\{1,2,3\}$.
As expected, $P(n|n)$ monotonically increases with $\eta$ such that at perfect detection efficiency $\eta=1$ for the number-resolving detector, it is certain that the $n$ photons detected arose from the input state  $\ket{n}_{a'}$.
As discussed above, state of the art number-resolving TES have efficiencies exceeding $87\%$ \cite{Schmidt:2018}. This implies a probability of $P(2|2)=0.71$ to detect the FS state $\ket{2}_{a'}$ from a strong TMSS source
($r=1.5\leftrightarrow -13$dB of squeezing) 
at $\eta=0.87$, resulting in the output state $\ket{2}_{b'}\to\ket{2}_{a}$ that can subsequently be mixed with the CS $\ket{\beta}_b$ to observe the CNL and PNC described above. For $\ket{3}_{b'}\to\ket{3}_{a}$ the probability drops to 
$P(3|3)=0.63$ for the same values of $r$ and $\eta$.
Bright SPDC sources in combination with photon number-resolving detectors have been used recently to experimentally realize multiphoton subtracted TMSS states by Maga\~{n}a-Loaiza \tit{et al}.
(see Fig.(2) of \cite{Magana-Loaiza:2019}) up to $10$ photons, based on theoretical proposals of Carranza and Gerry \cite{Carranza:2012}.

Since the extended HOM effects discussed in this work, CNL and PNC, depend on the properties of the beam splitter coefficients for non-classical FS input states, it is natural to inquire if such analogue effects are also present for a collection of atoms.
The answer is affirmative, and this analogy is discussed  in \App{app:HOM:CNL:Atoms} through the use of the Schwinger representation of $su(2)$, and the formal correspondence between a pair of dual mode Fock basis states $\ket{n,m}_{ab}$ with with angular momentum states $\ket{J,M}$ often used to describe a collection of atoms.
 
\section{Conclusion}\label{sec:Conclusion}
In this work we have shown that the parity of a nonclassical state of light has a dominant influence on the interference effects at a beam splitter, irrespective of the state it is mixed with at the other input port.
The parity of the nonclassical input carves deep valleys in the output joint number distribution.
A limiting case of our analysis is the Hong-Ou-Mandel effect, but we find dramatic additional richness in the interferences beyond this. This counter-intuitive influence of even a single photon to control the output of a beam splitter illuminated by any field, e.g. coherent or even a noisy thermal field, demonstrates the extraordinary power of non-classicality.  We explain the origin of these effects and explore prospects for observing them with currently available number-resolving detectors.

The extension of the HOM effect arises from the inherent ``filtering" property of a 50:50 BS when acting on non-classical states.
We have shown that if the $a$-mode of the BS contains only odd numbers of photons (an odd parity state), a central nodal line (CNL) exists for the output  joint probability distribution $P(m_a, m_b,\theta=\pi/2)$ for measuring $m_a$ photons in the $a$-mode and  $m_b$ photons in the $b$-mode, \tit{independent of the input state in the $b$-mode, pure or mixed}. In the simplest case of a single photon FS/FS input state $\ket{1,1}_{ab}$, this result reduces to the well known Hong-Ou-Mandel effect for the destructive interference of the 
quantum amplitude for the output $\ket{1,1}_{ab}$ state of a 50:50 BS.
We have shown that this CNL results from the \tit{intrinsic} property of the BS itself acting on non-classical Fock states (FS) of odd number.
This produces zeros in the angular portion of the beam splitter coefficients which dictate the mixing of the 
non-classical basis photon-pair FS  $\ket{n,m}_{ab}$ 
(for which arbitrary $a-$ and $b$-mode input states can be expanded)
such that $m_a+m_b=n+m$.

We have further shown that for a 50:50 BS there always exists  off-diagonal pseudo-nodal curves (PNC) upon which numerous sets of non-contiguous zeros lie, which carve out ``valleys" (local minima) in the output photon number distribution.
These PNC lie symmetrically placed about the diagonal symmetry line $m_a=m_b$ 
of the output joint photon number distribution
and exist for any input FS $\ket{n>0}_a$ entering the $a$ mode, with $n$ even or odd. For the case of a FS/CS input $\ket{n}_a\ket{\beta}_b$ we have provided explicit analytic solutions 
in parametric form $\big(m_a(k), m_b(k)\big)$ in terms of another integer $k\in\mathbb{Z}$. 
That the results presented in this work are universal has been explicitly demonstrated for various combinations of odd parity input states containing odd number of photons entering the $a$ mode mixed with arbitrary states entering the $b$ mode.
In addition, we have explored the existence of the PNC for a BS of arbitrary reflectivity, in particular $\theta=\pi/3$, and show that they must disappear as one approaches the extreme limits of 0 or 100\% BS transmittance. 

With today's state of the art photon-number resolving detectors (TES) using principal component analysis to histogram the energy bins for low number of photons (up to roughly $5$ reliably, possibly $10-20$ in some special cases, with high detection efficiencies) experiments to verify the predicted CNL and PNC should be experimentally feasible, even in the presence of a small amount of noise.

There is another feature of the output state joint photon number distribution that was first highlighted in the BMG paper \cite{BMG:2012} (where the CNL was noted and briefly discussed)
having to do with the relative heights of the constructive interference fringes appearing when mixing coherent light with increasing numbers of $n$ photons.  
As evident from \Fig{fig:FSCS:FSTS:plots:n0:1}  and \Fig{fig:FSCS:FSTS:plots:n2:3}, 
increasing the number of photons in the Fock state initially occupying the a mode tends towards pushing the peaks towards the two axes, creating a U-shape distribution along the anti-diagonal 
that becomes more pronounced as $n$ is taken to be larger.  
This is similar to the familiar distribution one gets when mixing twin Fock states at a balanced beam splitter to produce the arcsine states  \cite{Campos:1989,Gerry_Mimih:2010}.
Such a distribution, owing to its similarity to the well-known N00N state \cite{Dowling:2008}, has been known to be conducive to enhanced phase sensitivity beyond the standard quantum limit in interferometric measurements \cite{Campos:2003}.
The structure of this anti-diagonal U-shaped distribution in light of the extended HOM effect
will be investigated further in a future paper.

As a potential application, 
the results obtained in this work can be used in quantum key distribution protocols. If Alice and Bob decide to exchange qubits using nonorthogonal quantum states with different parity (i.e., odd or even photon number), the ability to produce joint photon distribution using number resolving detectors can help detect the presence of an eavesdropper. In principle, when using multi-photon states, an eavesdropper can gain sufficient information by carrying out photon-number splitting attack through quantum non-demolition measurements.  Eve separates the photons using an adjustable beamsplitter thereby sending a portion of the photons to Bob and storing the rest with her. However, because Bob only detects Eve by examining the detection rate, Eve’s presence can go unnoticed unless he uses methods, such as a decoy state technique, to overcome a photon-number-splitting attack. However, investigating the photon statistics on his end can be used to reveal the presence of Eve. Suppose Alice and Bob agree ahead of time that when they are ready to check for the presence of Eve, Alice will send a state of $n$ photons. As described, the joint photon distribution will exhibit $n$ PNC if $n$ is even and $n-1$ PNC plus one CNL if $n$ is odd. If Eve collects some of the photons transmitted to Bob, the number of PNC or PNC and CNL will change, revealing the presence of the eavesdropper. If Eve tries to detect photons and then transmit a copy, she will send an incorrect number of photons unless she has a perfect detector. 
This method requires that the same state must be transmitted enough times to build up the joint photon number distribution. 


Through the last three decades many researchers have hinted at the existence of features analogous to the CNL for special specific input state cases (typically involving a single photon in the $a$ mode and some interesting state in the $b$ mode). 
However, very little, if any, attention has been paid to the existence and origin of the PNC.
This work represents a comprehensive amalgamation of such past observations, plus a systematic explanation for their occurrence, as an intrinsic property of the 50:50 BS itself acting on non-classical states. Our results reduce to the well-known HOM effect in the limiting case of two single photon inputs to the BS, and hence can be rightfully deemed a 
general extension of the HOM effect.


\begin{acknowledgments}
PMA would like to acknowledge support of this work from
the Air Force Office of Scientific Research (AFOSR).
RJB would like to thank the National Academy of Sciences for support of this work through 
a National Research Council fellowship.
CCG would like to acknowledge support for this work was provided by
the Air Force Research Laboratory (AFRL) Visiting Faculty Fellowship Program (VFRP) under grant \#FA8750-20-3-1003.
The authors would like to thank M.L. Fanto and C.C. Tison for useful experimental discussions and considerations.
Any opinions, findings and conclusions or recommendations
expressed in this material are those of the author(s) and do not
necessarily reflect the views of Air Force Research Laboratory or the U.S. Navy.
\end{acknowledgments}

\nocite{apsrev41Control}
\bibliographystyle{apsrev4-1}
\bibliography{rr_losses_refs}

\begin{thebibliography}{41}
\bibitem [{Red()}]{Reduce:comment}%
  \BibitemOpen
\bibfield  {journal} {  }\href@noop {} {\bibinfo  {journal} {\ttt{Reduce} is
  \tit{Mathematica}'s routine to solve Diophantine equations. It should be
  noted that it does not return all solutions, and for the case of a 50:50 BS
  it only returned the lower branch solutions of
  $g(m_a,m_b,\theta=\pi/2|n=2,3)=0$ in
  \Fig{fig:gmambpidiv2n2n3_BruteForceOnlyn2_mambofk} and
  \Fig{fig:gmambpidiv3n2_analytic_solns_mambofk}(left). For potential solutions
  to $g(m_a,m_b,\theta=\pi/3|n=3)=0$
  \Fig{fig:gmambpidiv3n3_analytic_solns_mambofk}, \ttt{Reduce} returned
  non-polynomial solutions as roots of cubic equations $m_b = m_b(m_a)$}\
  }\BibitemShut {NoStop}%
\end{thebibliography}

\appendix
\clearpage
\newpage
 \section{Action of the BS on $\boldsymbol{\ket{n,m}_{ab}}$}\label{app:BS}
\begin{figure}[ht]
\begin{center}
\includegraphics[width=3.5in,height=1.55in]{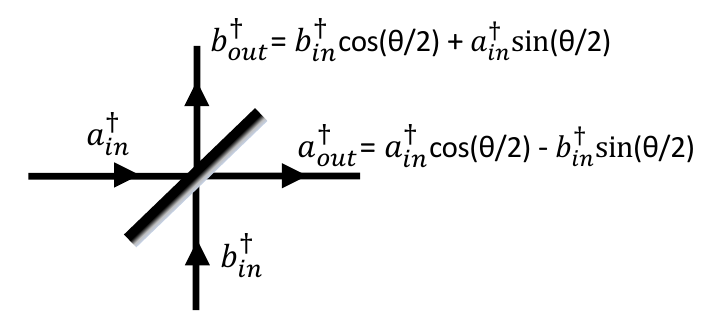}
\caption{\label{fig:BS:v1:app} Two optical modes $a$ and $b$ mixing on a BS of transmissivity $T = \cos^2(\theta/2)$. 
Here, the bottom of the BS imparts a $\pi$ phase shift of $-1$ upon reflection.
}
\end{center}
\end{figure}

 We need to know how an ideal, lossless BS acts on an arbitrary 
 two-mode FS input basis state $\ket{n,m}_{ab}$ presented at its two input ports.
 Let us define the BS transformation (Hamiltonian) on two modes $a$ and $b$ as $BS=(\theta/2)\,(a\,b^\dag + a^\dag\,b)$.
 Here $T \equiv\cos^2(\theta/2)$ is the BS transmissivity, and 
 $R =(1-T) = \sin^2(\theta/2)$ is the BS reflectivity  such that $R+T=1$, as shown in \Fig{fig:BS:v1:app}. 
 (Note: we call the quantities $\sin(\theta/2)$ and $\cos(\theta/2)$ reflection and transmission \tit{coefficients}).
 The factor of $1/2$ in the argument $\theta/2$ is introduced so that $\theta=\pi/2$ represents a 50:50 BS.
%

\subsection{Conventions and operator transformations}\label{app:subsec:conventions}
 The action of the BS on an arbitrary Fock pair basis state 
$\ket{n}_a\ket{m}_b \equiv \dfrac{(a^\dag)^n}{\sqrt{n!}}\,\dfrac{(b^\dag)^m}{\sqrt{m!}}\,\ket{0}_a\ket{0}_b$ is straightforwardly computed 
(see Chapter 5 of Agarwal \textit{Quantum Optics} \cite{Agarwal:2013})
by applying the BS transformation to the last expression, and expanding out terms using the binomial theorem (since $a^\dag$ and $b^\dag$ commute).

Here we take the fundamental transformation of the annihilation operators by the unitary $U(\theta)\equiv U_{BS}(\theta)$ to be
\bea{out:in:and:in:out:app}
\vec{a}_{out}&=&
\left[
\begin{array}{c}
a_{out} \\
b_{out}
\end{array}
\right] = 
U(\theta)
\left[
\begin{array}{c}
a_{in} \\
b_{in}
\end{array}
\right] 
U^\dag(\theta) \no
&=&
\left[
\begin{array}{cc}
\cos(\theta/2) & -\sin(\theta/2)\\
\sin(\theta/2) & \cos(\theta/2)
\end{array}
\right]
\,
\left[
\begin{array}{c}
a_{in} \\
b_{in}
\end{array}
\right] 
\equiv
S^T_{BS}(\theta)\, \vec{a}_{in}, \qquad\;
\eea
where we have defined the matrix $S^T_{BS}(\theta)$ by the $2\times 2$ rotation matrix in \Eq{out:in:and:in:out:app}.

Note that if we write the BS transformation $S_{BS}$ of the \textit{out} operators in terms of the \textit{in} operators as
$\vec{a}^\dag_{out} = S^T_{BS}\, \vec{a}^\dag_{in}$
then to transform an input state such as $\ket{1}_a\,\ket{0}_b = a^\dag_{in}\ket{0}_a\,\ket{0}_b$, we need to
write the \textit{in} operators in terms of the \textit{out} operators using the \textit{transpose} transformation 
$S_{BS}$ as
$\vec{a}^\dag_{in} = S_{BS}\, \vec{a}^\dag_{out}$  \cite{Skaar:2004}  via
\bea{out:in:and:in:out}
\hspace{-0.2in}
\vec{a}^\dag_{out}&=&
\left[
\begin{array}{c}
a^\dag_{out} \\
b^\dag_{out}
\end{array}
\right] = 
\left[
\begin{array}{cc}
\cos(\theta/2) & -\sin(\theta/2)\\
\sin(\theta/2) & \cos(\theta/2)
\end{array}
\right]
\,
\left[
\begin{array}{c}
a^\dag_{in} \\
b^\dag_{in}
\end{array}
\right] 
\equiv
S^T_{BS}\, \vec{a}^\dag_{in}, \quad\;\; \label{out:in:and:in:out:line1}\\
\hspace{-0.2in}
\Rightarrow \vec{a}^\dag_{in} &=& 
\left[
\begin{array}{c}
a^\dag_{in} \\
b^\dag_{in}
\end{array}
\right] 
=
\left[
\begin{array}{cc}
\cos(\theta/2) & \sin(\theta/2)\\
-\sin(\theta/2) & \cos(\theta/2)
\end{array}
\right]
\,
\left[
\begin{array}{c}
a^\dag_{out} \\
b^\dag_{out}
\end{array}
\right] 
\equiv S_{BS}\, \vec{a}^\dag_{out}, \quad\;\;\;\; \label{out:in:and:in:out:line2}\\
\hspace{-0.2in}
&{}& \no
\hspace{-0.2in}
&\Rightarrow&
U(\theta) \,\vec{a}^\dag_{in}\,U^\dag(\theta) =S_{BS}\,\vec{a}^\dag_{out}.  \label{out:in:and:in:out:line3}
\eea
Thus, for example
\bea{BS:example}
\hspace{-0.25in}
\ket{\psi_{in}}_{ab}&=&\ket{1}_a\ket{0}_b = a^\dag_{in}\ket{0}_a\,\ket{0}_b,\no
\hspace{-0.25in}
\Rightarrow \ket{\psi_{out}}_{ab} &=& U(\theta)  a^\dag_{in}\ket{0}_a\,\ket{0}_b,\no
\hspace{-0.25in}
&=& \left( U(\theta)  a^\dag_{in} U^\dag(\theta)\right)\, U(\theta)\ket{0}_a\,\ket{0}_b,\no
\hspace{-0.25in}
&=& \left( S_{BS}\, \vec{a}^\dag_{out}\right)\,\ket{0}_a\,\ket{0}_b, \; (\trm{using}\, U(\theta)\ket{0}_a\,\ket{0}_b = \ket{0}_a\,\ket{0}_b), \no
\hspace{-0.25in}
&=& \left(\cos(\theta/2)\,a^\dag_{out} + \sin(\theta/2)\,b^\dag_{out}\right)\, \ket{0}_a\,\ket{0}_b,\no
\hspace{-0.25in}
&=& \cos(\theta/2)\,\ket{1}_a\,\ket{0}_b + \sin(\theta/2)\ket{0}_a\,\ket{1}_b. 
\eea
We can now drop all the  \textit{in, out} labels and just remember to use the transformation 
$S_{BS}$ 
when applying the BS transformation to creation operators when transforming
 from input states to output states, i.e.
 \bsub
\bea{BS:opr:transf} 
\hspace{-0.25in}
\ket{\psi_{in}}_{ab} &\equiv& f(\vec{a}^\dag) \,\ket{0}_a\,\ket{0}_b, 
\;\trm{for some function $f$ of}\;
\vec{a}^\dag =
\left[
\begin{array}{c}
a^\dag \\
b^\dag
\end{array}
\right], \qquad\;\; \label{BS:opr:transf:line1} \\
\hspace{-0.25in}
\Rightarrow \ket{\psi_{out}}_{ab} &=& U(\theta) \, \ket{\psi_{in}}_{ab}, \label{BS:opr:transf:line2} \\
&=&  U(\theta) f(\vec{a}^\dag)\,U^\dag(\theta) \,U(\theta)\,\ket{0}_a\,\ket{0}_b \no
&=&   f(S_{BS}\vec{a}^\dag)\,\ket{0}_a\,\ket{0}_b, 
\quad \trm{using}\;\quad \vec{a}^\dag\overset{U(\theta)}{\longrightarrow} S_{BS}\,\vec{a}^\dag, \quad \\
&{}&
S_{BS} = 
\left[
\begin{array}{cc}
\cos(\theta/2) & \sin(\theta/2)\\
-\sin(\theta/2) & \cos(\theta/2)
\end{array}
\right]. \label{BS:opr:transf:line3}
\eea
\esub
\Eq{BS:opr:transf:line1}, \Eq{BS:opr:transf:line2}, and \Eq{BS:opr:transf:line3} are the primary BS operator transformation equations that we will employ throughout this work.

\subsection{BS transformation of the state $\ket{n,m}_{ab}$ and examples}
We are now interested in the BS transformed output state 
$ \ket{n,m}_{a b}^{(out)} = U(\theta)\,\ket{n,m}_{a b}$ 
when $n$ photons are present at input mode $a$, and  
$m$ photons are present at input mode $b$.
The derivation is easily carried out 
(see Agarwal \cite{Agarwal:2013}) 
with the results given below 
using $S_{BS}$ to transform an input state $\ket{n}_a\ket{m}_b$ to an output state, yielding
\bsub
\bea{BS:Agarwal}
 \ket{n,m}_{ab} &\to& \ket{n,m}^{(out)}_{ab}\equiv  \sum_{p=0}^{n+m}\, f^{(n,m)}_p \,\ket{p}_a\,\ket{n+m-p}_b, \label{BS:on:n:m} \\
%
 f^{(n,m)}_p &=& \sum_{q=0}^n  \sum_{q'=0}^m \delta_{p,q+q'}\, \binom{n}{q}\,\binom{m}{q'}\,
\sqrt{\dfrac{p!\,(n+m-p)!}{n!\,m!}}\, \times
\no
&{}& \left(-1\right)^{q'}\,
   \left(\cos(\theta/2)\right)^{m+q-q'}
\, \left(\sin(\theta/2)\right)^{n-q+q'}, \qquad\quad  \label{BS:Agarwal:fnmp}
\eea 
\esub
\footnote{
Note: Agarwal uses 
$S_{BS} = 
\tiny{\left[
\begin{array}{cc}
\cos(\theta) & i\,\sin(\theta)\\
i\,\sin(\theta) & \cos(\theta)
\end{array}
\right]}$ 
which results in 
$\cos(\theta/2)\to \cos(\theta)$, 
$\sin(\theta/2)\to i\,\sin(\theta)$,
and the absence of the factor of $(-1)^{q'}$
in \Eq{BS:Agarwal:fnmp}. 
The crucial minus signs in $g(m_a,m_b,\theta)$ then appear from factors of $(i^2)$ raised to various powers.
}.
Note that the delta function $\delta_{p,q+q'}$ ensures that the BS mixes the original input state $\ket{n}_a\ket{m}_b$ only amongst the
$n+m+1$ states of total photon number $n+m$ of the form 
$\{ \ket{0}_a\ket{n+m}_b,\, \ket{1}_a\ket{n+m-1}_b, \ldots  \ket{n}_a\ket{m}_b,\ldots,   \ket{n+m-1}_a\ket{1}_b,\  \ket{n+m}_a\ket{0}_b\}$.
The (real) BS coefficients $f^{(n,m)}_p$ are easily worked out by hand by considering states  
$\ket{n}_a\ket{m}_b$ up to $n+m=2$ at the input ports of BS, namely: 
\bwt
%
%
\bsub
\bea{fp:n:0:1}
n+m=0: &{} \no
\ket{0}_a\ket{0}_b &\to& \ket{0}_a\ket{0}_b \Rightarrow  f^{(0,0)}_0 = 1, \\
n+m=1: &{} \no
\ket{0}_a\ket{1}_b &\to&  [\cos(\theta/2)\,b^\dag - \sin(\theta/2)\,a^\dag ]\ket{0}_a\ket{0}_b, \no
                   					          &=&  \cos(\theta/2) \ket{0}_a\ket{1}_b - \sin(\theta/2) \ket{1}_a\ket{0}_b, 
					                           \Rightarrow   \left\{ \begin{array}{ccr}  f^{(0,1)}_0 &=& \cos(\theta/2) \\
					                                                                                       f^{(0,1)}_1 &=& -\sin(\theta/2) 
					                                                          \end{array}  
					                                                 \right. \label{fnm01p:app}\\ 
\ket{1}_a\ket{0}_b &\to&   [\cos(\theta/2)\,a^\dag + \sin(\theta/2)\,b^\dag ]\ket{0}_a\ket{0}_b, \no
                   					          &=&   \sin(\theta/2) \ket{0}_a\ket{1}_b + \cos(\theta/2) \ket{1}_a\ket{0}_b, 
					                           \Rightarrow   \left\{ \begin{array}{ccr}  f^{(1,0)}_0 &=& \sin(\theta/2), \\
					                                                                                      f^{(1,0)}_1 &=&   \cos(\theta/2), 
					                                                          \end{array}  
					                                                 \right. 
\eea
\bea{fp:n:2}
n+m=2: &{} \no
\ket{0}_a\ket{2}_b &=&  \dfrac{1}{\sqrt{2}}\, b^{\dag 2}\,\ket{0}_a\ket{0}_b \to  \dfrac{1}{\sqrt{2}}\,[\cos(\theta/2)\,b^\dag - \sin(\theta/2)\,a^\dag ]^2\ket{0}_a\ket{0}_b, \no
                   					          &=&  \cos^2(\theta/2) \ket{0}_a\ket{2}_b -\sin(\theta) \ket{1}_a\ket{1}_b +  \sin^2(\theta/2) \ket{2}_a\ket{0}_b, 
					                           \Rightarrow   \left\{ \begin{array}{ccl}  f^{(0,2)}_0 &=& \cos^2(\theta/2), \\
					                                                                                       f^{(0,2)}_1 &=& -\frac{1}{\sqrt{2}}\,\sin(\theta),  \\
					                                                                                       f^{(0,2)}_2  &=& \sin^2(\theta/2), 
					                                                          \end{array}  
					                                                 \right. \\ 			                                                   
\ket{1}_a\ket{1}_b   &=& a^\dag\,b^\dag\,\ket{0}_a\ket{0}_b \to [\cos(\theta/2)\,a^\dag + \sin(\theta/2)\,b^\dag ]\,[\cos(\theta/2)\,b^\dag - \sin(\theta/2)\,a^\dag ]\ket{0}_a\ket{0}_b^\dag, \no
                   					          &=&  \sqrt{2}\,\sin(\theta/2)\,\cos(\theta/2) \ket{0}_a\ket{2}_b 
					                                    [\cos^2(\theta/2) - \sin^2(\theta/2)] \ket{1}_a\ket{1}_b , \no
					                                    & & \hspace{1.75in} - \sqrt{2}\sin(\theta/2)\,\cos(\theta/2) \ket{2}_a\ket{0}_b, 
					                           \Rightarrow   \left\{ \begin{array}{ccl}  f^{(1,1)}_0 &=& \frac{1}{\sqrt{2}} \sin(\theta), \\
					                                                                                       f^{(1,1)}_1 &=&                             \cos(\theta),  \\
					                                                                                       f^{(1,1)}_2  &=& -\frac{1}{\sqrt{2}} \sin(\theta), 
					                                                          \end{array}  
					                                                 \right. \\ 
\ket{2}_a\ket{0}_b  &=&  \dfrac{1}{\sqrt{2}}\, a^{\dag 2}\,\ket{0}_a\ket{0}_b \to \dfrac{1}{\sqrt{2}}\, [\cos(\theta/2)\,a^\dag + \sin(\theta/2)\,b^\dag ]^2\ket{0}_a\ket{0}_b, \no
                   					          &=&  \sin^2(\theta/2) \ket{0}_a\ket{2}_b +\sin(\theta) \ket{1}_a\ket{1}_b +  \cos^2(\theta/2) \ket{2}_a\ket{0}_b, 
					                           \Rightarrow   \left\{ \begin{array}{ccl}  f^{(2,0)}_0 &=& \sin^2(\theta/2), \\
					                                                                                       f^{(2,0)}_1 &=& \frac{1}{\sqrt{2}}\,\sin(\theta),  \\
					                                                                                       f^{(2,0)}_2  &=& \cos^2(\theta/2). 
					                                                          \end{array}  
					                                                 \right. \qquad					                                                 
\eea
\esub
%
%
%
\ewt

We can check that the general formula  \Eq{BS:Agarwal:fnmp} gives the correct expressions (and signs) by examining
$f^{(0,1)}_p$ worked out above in  \Eq{fnm01p:app}.
From \Eq{BS:Agarwal:fnmp} with $n=0, m=1$  (which implies $q=0$, $q' = p-q = p$) we have
\bea{fnm01p:from:formula}
\hspace{-.5in}
f^{(0,1)}_p &=& \binom{1}{p}\,\sqrt{p!\,(1-p)!} \,(-1)^p \, (\cos(\theta/2))^{1-p}\,(\sin(\theta/2))^p, \qquad\;\; \\
\hspace{-.5in}
\Rightarrow
f^{(0,1)}_0 &=& \binom{1}{0}\,\sqrt{0!\,(1-0)!} \,(-1)^0 \, (\cos(\theta/2))^{1-0}\,(\sin(\theta/2))^0\no &=& \cos(\theta/2), \no
\hspace{-.5in}
\Rightarrow
f^{(0,1)}_1 &=& \binom{1}{1}\,\sqrt{1!\,(1-1)!} \,(-1)^1 \, (\cos(\theta/2))^{1-1}\,(\sin(\theta/2))^1\no &=& -\sin(\theta/2), \nonumber
\eea
which agrees with the hand calculation of  \Eq{fnm01p:app}.

Note: for each $(n,m)$ we have $\sum_{p=0}^{n+m} | f^{(n,m)}_p|^2 = 1$, which just indicates that the BS transformation is unitary.
Note  that the $f^{(n,m)}_p(\theta)$  are just the Wigner rotation coefficients for the representation of a system with spin $J = (n+m)/2$ in the angular momentum basis
$\ket{J,M}$ with $2 J + 1 = n+m+1$ states $M\in\{-J, -J+1,\ldots,J\}$ where $M(p) = -J + p\, (2 J) / (n+m)$ for $p\in\{0,\ldots,n+m\}$
(see also Appendix A and B of \cite{BAG_Parity:2021}).

\subsection{Derivation of the BS coefficients $\boldsymbol{ f^{(n,m)}_p(\theta)}$ \Eq{BS:Agarwal:fnmp}, and their hyper-binomial form}\label{app:subsec:BS:derivation}
The derivation of the BS coefficients (BSCs) $f^{(n,m)}_p(\theta)$ in \Eq{BS:Agarwal:fnmp} is a straightforward exercise 
(see Agarwal, Chapter 5 \cite{Agarwal:2013}).
Consider the dual Fock input state $\ket{n}_a\ket{m}_b$ entering the BS.
Using the BS transformation rules and the binomial expansion we have
\bwt
\small
\bsub
\bea{BSC:derivation:app}
\hspace{-0.9in}
\ket{n}_a\ket{m}_b &=& \dfrac{(a^\dag)^n}{\sqrt{n!}} \dfrac{(b^\dag)^m}{\sqrt{m!}} \ket{0,0}_{ab}, \no
 &\overset{U(\theta)}{\longrightarrow}& 
 \dfrac{(a^\dag \cos(\theta) + b^\dag\,\sin(\theta))^n}{\sqrt{n!}} \dfrac{(b^\dag \cos(\theta) - a^\dag\,\sin(\theta))^m}{\sqrt{m!}} \ket{0,0}_{ab},\no
&=& 
\frac{1}{\sqrt{n! m!}}\,
\sum_{q=0}^n   \binom{n}{q}(\cos(\theta/2)a^\dag)^{q}(\sin(\theta/2)b^\dag)^{n-q}
\sum_{q'=0}^m  \binom{m}{q'}(\cos(\theta/2)b^\dag)^{m-q'}(-\sin(\theta/2)a^\dag)^{q'} \ket{0,0}_{ab},\no
&=&
\sum_{q=0}^n \sum_{q'=0}^m  \binom{n}{q} \binom{m}{q'} (-1)^{q'}(\cos(\theta/2))^{m+(q-q')} (\sin(\theta/2))^{n-(q-q')}
(a^\dag)^{q+q'} (b^\dag)^{n+m-(q+q')} \ket{0,0}_{ab}, \no
&=&
\sum_{q=0}^n \sum_{q'=0}^m 
\binom{n}{q} \binom{m}{q'}\sqrt{\frac{(q+q')(n+m-(q+q'))}{n! m!}}
 (-1)^{q'}(\cos(\theta/2))^{m+(q-q')} (\sin(\theta/2))^{n-(q-q')}
\ket{q+q', n+m-(q+q')}_{ab}, \no
&\equiv&
\sum_{p=0}^{n+m}\sum_{q=0}^n \sum_{q'=0}^m \delta_{p,q+q'}
\binom{n}{q} \binom{m}{q'}\sqrt{\frac{(q+q')(n+m-(q+q'))}{n! m!}}
 (-1)^{q'}(\cos(\theta/2))^{m+(q-q')} (\sin(\theta/2))^{n-(q-q')}
\ket{p, n+m-p}_{ab}, \no
&\equiv& \sum_{p=0}^{n+m}\, f^{(n,m)}_p \,\ket{p}_a\,\ket{n+m-p}_b, \label{BSC:derivation:line:1:app} \\
\hspace{-0.9in}
f^{(n,m)}_p(\theta) &\equiv&
 \sum_{q=0}^n  \sum_{q'=0}^m \delta_{p,q+q'}\, 
 \binom{n}{q}\,\binom{m}{q'}\,
\sqrt{\dfrac{p!\,(n+m-p)!}{n!\,m!}}\,
\left(-1\right)^{q'}\,
   \left(\cos(\theta/2)\right)^{m+(q-q')}\,
   \left(\sin(\theta/2)\right)^{n-(q-q')}, \label{BSC:derivation:line:2:app}\\
&=&   
\sqrt{\dfrac{p!\,(n+m-p)!}{n!\,m!}}\,
\left(-1\right)^{p}\,
 \sum_{q=0}^n \binom{n}{q}\,\binom{m}{q'}\,\left(-1\right)^{q}\,
    \left(\cos(\theta/2)\right)^{m+(2q-p)}\,
   \left(\sin(\theta/2)\right)^{n-(2q-p)}, \label{BSC:derivation:line:3:app}\\
&=&   
\sqrt{\dfrac{p!\,(n+m-p)!}{n!\,m!}}\,
 \left(\cos(\theta/2)\right)^{m-p}
   \left(\sin(\theta/2)\right)^{n+p}\,
   \left(-1\right)^{p}\,
 \sum_{q=0}^n \binom{n}{q}\,\binom{m}{p-q}\,\left(\frac{-1}{\tan^2(\theta/2)}\right)^q, \label{BSC:derivation:line:4:app} \\
 &\equiv&
\sqrt{\dfrac{p!\,(n+m-p)!}{n!\,m!}}\,
 \left(\cos(\theta/2)\right)^{m-p}
   \left(\sin(\theta/2)\right)^{n+p}\,
    \left(-1\right)^{p}\,
   \binom{m}{p} \, _2F_1\Big(-n,-p; 1+m-p; \frac{-1}{\tan^2(\theta/2)}\Big), \label{BSC:derivation:line:5:app}
\eea
\esub
\normalsize
\ewt
where in the last expression \Eq{BSC:derivation:line:5:app} we have formally performed the binomial sum in terms of the hypergeometric function $_2F_1(a,b;c;z)$. 
Note that in the line before \Eq{BSC:derivation:line:1:app} we have introduced the new summation variable $p\in\{0,n+m\}$ 
and the ``boundary condition" that $p\equiv q+q'$ by including $\sum_{p=0}^{n+m} \delta_{p,q+q'}$ to ensure that the 
FS $\ket{p}_a$ and  $\ket{n+m-p}_b$ stay ``in bounds." In \Eq{BSC:derivation:line:4:app} we have performed the sum over $q'$, 
yielding $q'=p-q$ (and noting that if the $q$ index goes ``out of bounds" one obtains 0 from the binomial coefficients, i.e. $\binom{m}{p-q}=0$ for $p-q<0$, since $1/(-k)! = 0$ for integer $k\in\mathbb{Z}_{\ge 0}$), and then pull out all non-$q$ dependent terms from the sum over $q$.  

We now put \Eq{BSC:derivation:line:4:app} in one last final \tit{hyper-binomial} form, that will be conducive when we compute probabilities of $m_a, m_b$ photons in mode $a, b$ respectively. Let us consider the binomial term 
$\binom{m}{p-q} = \tfrac{m!}{(p-q)!(m-p+q)!} \equiv \tfrac{m!}{(p-q)!((m +n-p)- (n-q))!}$ in the summation in \Eq{BSC:derivation:line:4:app}. Now using the definition of the \tit{falling factorial} $(x)_k = \tfrac{x!}{(x-k)!}$ we can replace the factorial terms in the denominator using
 $\tfrac{1}{(x-k)!}= \tfrac{(x)_k}{x!}$, so that  $\binom{m}{p-q} = \tfrac{m!\,(p)_q\,(m+n-p)_{n-q}}{p! (m+n-p)!}$. 
 In anticipation of measurement results to come, we also write $m! \equiv ((m+n)-n)! = \tfrac{(m+n)!}{(m+n)_n}$. 
 This yields the final desired form
 $\binom{m}{p-q} = \binom{m+n}{p}\,\tfrac{(p)_q\,(m+n-p)_{n-q}}{(m+n)_{n}}$.
 Lastly, by multiplying the summand by $1=\tfrac{\tan^{2n}(\theta/2)}{\tan^{2n}(\theta/2)}$, 
 and pulling out all non-$q$ terms from the summation, 
 we can write \Eq{BSC:derivation:line:4:app} finally as 
 %
%
\bwt
\bea{fnmp:final}
\hspace{-0.65in}
f^{(n,m)}_p(\theta) 
&=& 
\frac{(-1)^p}{\sqrt{n!\, (m+n)_n}}\,\sqrt{\binom{m+n}{p}}\, 
\big(\cos(\theta/2) \big)^{m-p}\,\big(\sin(\theta/2) \big)^{p-n}
\sum_{q=0}^{n} \binom{n}{q}\,(-1)^q\, (p)_q\, \big(\cos^2(\theta/2) \big)^q \,(m+n-p)_{(n-q)}\,\big(\sin^2(\theta/2) \big)^{n-q}, 
\label{fnmp:final:line1} \no
\hspace{-0.65in}
&\equiv& 
\frac{(-1)^{p+n}}{\sqrt{n!\, (m+n)_n}}\,\sqrt{\binom{m+n}{p}}\, 
\big(\cos(\theta/2) \big)^{m-p}\,\big(\sin(\theta/2) \big)^{p-n}
\sum_{q=0}^{n} \binom{n}{q}\,(-1)^q\, (p)_{n-q}\, \big(\cos^2(\theta/2) \big)^{n-q} \,(m+n-p)_{q}\,\big(\sin^2(\theta/2) \big)^{q},   \no
&\equiv& 
\frac{(-1)^{p+n}}{\sqrt{n!\, (m+n)_n}}\,\sqrt{\binom{m+n}{p}}\, 
\big(\cos(\theta/2) \big)^{m-p}\,\big(\sin(\theta/2) \big)^{p-n}\,g^{(n,m)}_p(\theta). \label{fnmp:final:line2}
\eea
%
%
\ewt
%
In the middle expression we have used $\binom{n}{q} = \binom{n}{n-q}$ and changed summation variables in the first expression to $q'=n-q$ and used
$(-1)^q = (-1)^{(n-q')} = (-1)^n\,(-1)^{q'}$. Relabeling $q'\to q$ again, then simply switches $q\leftrightarrow (n-q)$ in \Eq{fnmp:final:line1} and introduces an over factor of $(-1)^n$ which can be pulled out of the summation over $q$.

We see that the zeros of $f^{(n,m)}_p(\theta)$ are governed by the summation over $q$ term $g^{(n,m)}_p(\theta)$ 
in \Eq{fnmp:final:line2} which we have defined as
\bwt
\be{gnmptheta}
g^{(n,m)}_p(\theta) \equiv \sum_{q=0}^{n} \binom{n}{q}\,(-1)^q\, (p)_{n-q}\,
 \big(\cos^2(\theta/2) \big)^{n-q} \,(m+n-p)_{q}\,\big(\sin^2(\theta/2) \big)^{q}.
\ee
\ewt

Since the state exiting the BS with a dual FS input $\ket{n,m}_{ab}$ is given by
$\ket{n,m}^{(out)}_{ab}=\sum_{p=0}^{n+m}\,~f^{(n,m)}_p \,\ket{p}_a\,\ket{n+m-p}_b$, the amplitude to be in the output state
$\ket{m_a, m_b}_{ab}$ is given by
$ {}_{ab}\IP{m_a,m_b}{n,m}^{(out)}_{ab} = f^{(n,m)}_p(\theta)\, \delta_{p,m_a}\, \delta_{m,m_a+m_b-n}$.
Note that the particular form of \Eq{gnmptheta} was chosen in anticipation of the above delta functions which set
$p\to m_a$ and $m+n-p\to (m_a+m_b-n) + n - m_a) = m_b$.

Thus, the joint photon number probability $P(m_a, m_b, \theta)$ is proportional to 
$\left(g^{(n,m=m_a,m_b)}_{p=m_a}(\theta)\right)^2 \equiv \big(g(m_a,m_b,\theta|n)\big)^2$.
From \Eq{fnmp:final:line2} and \Eq{gnmptheta} we write the beam splitter coefficient after measurement as
%
\bwt
\bsub
\bea{fmambthetan}
\hspace{-0.75in}
f^{(n,m=m_a+m_b-n)}_{p=m_a}(\theta) &=& 
\frac{(-1)^{m_a+n}}{\sqrt{n!\, (m_a+m_b)_n}}\,\sqrt{\binom{m_a+m_b}{m_a}}\, 
\big(\cos(\theta/2) \big)^{m_b-n}\,\big(\sin(\theta/2) \big)^{m_a-n}\,g^{(n,m_a+m_b-n)}_{p=m_a}(\theta), \label{fmambthetan:line1}\\
\hspace{-0.75in}
g^{(n,m=m_a,m_b)}_{p=m_a}(\theta) &=&
\sum_{q=0}^{n} \binom{n}{q}\,(-1)^q\, (m_a)_{n-q}\, \big(\cos^2(\theta/2) \big)^{n-q} \,(m_b)_{q}\,\big(\sin^2(\theta/2) \big)^{q}
\equiv g(m_a,m_b,\theta|n), \label{fmambthetan:line2} \\
&\equiv & (m_a)_n\, \left(\cos^{2}(\theta/2) \right)^n \, {}_2F_1(-m_b, -n; m_a+1-n;-\tan^2(\theta/2)). \label{fmambthetan:line3}
\eea
\esub
%
\ewt
 \Eq{fmambthetan:line1} and  \Eq{fmambthetan:line2} are the primary equations from which we deduce the main results of this work.
Note \Eq{fmambthetan:line3} defines the formal sum of \Eq{fmambthetan:line2}  in terms of the hypergeometric function
${}_2F_1(a,b;c;z)$ with $a$ and $b$ negative integers, which is well defined (see \cite{AbramowitzStegun:1972}),
but nonetheless not representable in any form of special function whose properties can be readily exploited, even at the 50:50 BS angle $\theta=\pi/2$.
However, in practice we find \Eq{fmambthetan:line2} more useful, and more intuitively appealing.
%
%
%
The crucial property of \Eq{fmambthetan:line2} is that one can always factor out $\cos(\theta)$ when $n$ is odd.
This non-trivial (and heretofore  overlooked) property is not readily apparent from \Eq{fmambthetan:line3} and properties of the hypergeometric function ${}_2F_1$, since the latter involves factors of 
$\cos^2(\theta/2)$ and $\tan^2(\theta/2)$ vs $\cos(\theta) = \cos^2(\theta/2)-\sin^2(\theta/2)$.
Note: if $n=odd$ ($even$), \Eq{fmambthetan:line2} contains and even (odd) number of terms. 
It is this particular ``near binomial" form of \Eq{fmambthetan:line2} 
that has allowed to deduce properties of FS,  and hence superposition of FS, entering the ports of the BS.

\clearpage
\newpage
\section{Proof of \Eq{gmmtheta:n:odd}} \label{app:Proof}
In this appendix we prove that we can always factor out $\cos(\theta)$
from $g(m_a,m_b,\theta|n)$ when $n$ is odd, which is the origin of the CNL in the extended HOM effect.
From \Eq{fmambthetan:line2} we have 
\bwt
\bsub
\bea{gmmthetaxy}
g(m_a,m_b,\theta|n)
&=&
\sum_{q=0}^{n} \binom{n}{q}\,(-1)^q\, 
(m_a)_{n-q}\, \big(\cos^2(\theta/2) \big)^{n-q} 
\,(m_b)_{q}\,\big(\sin^2(\theta/2) \big)^{q}, \label{gmmthetaxy:line1} \\
\overset{m_a=m_b=m'}{\Longrightarrow} 
g(m',m',\theta|n) &=&
 \sum_{q=0}^{n} \binom{n}{q}\,(-1)^q\, 
(m')_{n-q}\,(m')_{q}\, 
\big(\cos^2(\theta/2) \big)^{n-q}\big(\sin^2(\theta/2) \big)^{q}, \label{gmmthetaxy:line2} \\
&\equiv&
 \sum_{q=0}^{n} \binom{n}{q}\,(-1)^q\, 
(m')_{n-q}\,(m')_{q}\, 
x^{n-q}\,y^{q},  \label{gmmthetaxy:line3} 
\eea
\esub
\ewt
where for convenience we have defined $x=\cos^2(\theta/2)$ and $y=\sin^2(\theta/2)$.
The goal is to show that for $n$ odd, one can always factor out 
$x-y=\cos^2(\theta/2)-\sin^2(\theta/2)=\cos(\theta)\overset{\theta=\pi/2}{\longrightarrow}0$ for a 50:50 BS
(since the CNL is then given by $P(m',m',\theta|n)\propto g^2(m',m',\theta|n)\overset{\theta=\pi/2}{\longrightarrow}0$).

For $n$ odd, there are an even number of terms in \Eq{gmmthetaxy:line3}, the first $(n-1)/2$ of which have the same magnitude, but opposite sign of the latter $(n-1)/2$ terms. This allows us to write
\bwt
\bsub
\bea{gmmthetaxy:sym}
g(m',m',\theta|n)
&=&
 \sum_{q=0}^{(n-1)/2} \binom{n}{q}\,(-1)^q\, 
(m')_{n-q}\,(m')_{q}\, 
\left[
x^{n-q}\,y^{q}
-x^{q}\,y^{n-q}
\right], \label{gmmthetaxy:sym:line1} \\
&\equiv&
(x-y)\,\sum_{q=0}^{(n-1)/2} \binom{n}{q}\,(-1)^q\, 
(m')_{n-q}\,(m')_{q}\, 
x^{q}\,y^{q}
\left[
\sum_{k=1}^{n-2 q}
x^{(n-2q)-k}\,y^{k-1}  \label{gmmthetaxy:sym:line2}
\right],
\eea
\esub
\ewt
which is the proof of the main assertion. 
In moving from  \Eq{gmmthetaxy:sym:line1} to \Eq{gmmthetaxy:sym:line2} we have used the algebraic identities that for fixed $q\in\{0,1,\ldots,(n-1)/2\}$ with $n$ odd, and for any integer $r$ we have
\bsub
\bea{algebraic:identities}
\hspace{-0.6in}
x^{n-q}\,y^{q}
-x^{q}\,y^{n-q} &=& 
x^q\,y^q\,(x^{n-2q} - y^{n-2q}) \no
&=&
%
 x^q\,y^q\,(x-y)\,\sum_{k=1}^{n-2 q}
x^{(n-2q)-k}\,y^{k-1},\;\;  \label{algebraic:identities:1} \\
%
x^r - y^r &=& (x-y) \, \sum_{k=1}^{r} x^{r-k}\,y^{k-1}.  \label{algebraic:identities:2}
\eea
\esub
\Eq{algebraic:identities:2} follows simply by expanding out the terms on the right hand side and seeing that all terms cancel in pairs, except for the first and last unpaired terms $x^r$ and $-y^r$.
The first identity \Eq{algebraic:identities:1} is best illustrated by explicitly checking its validity on small values of $n$, and then proving by induction.
For example, using $n=5$, \Eq{algebraic:identities:1} yields 
\bsub
\begin{alignat}{3}
&x^{n-q}\,y^{q}&&- x^{q}\,y^{n-q} && \to n= 5,\quad q\in\{0,1,2\},   \\
&q=0:\; &&x^5 - y^5 &&= x^0\,y^0\,(x^5 - y^5) \\ 
&{}  &&{} &&=(x-y)\,(x^4 + x^3\,y + x^2\,y^2 + x\,y^3 + y^4),  \no
&q=1:\; &&x^4\,y - x\,y^4 &&=x^1\,y^1\,(x^3-y^3) \\ 
&{}  &&{} &&= x\,y\,(x-y)\,(x^2 +  x\,y + y^2),\no
&q=2:\; &&x^3\,y^2 - x^2\,y^3 && = x^2\,y^2\,(x^1-y^1), 
\end{alignat}
\esub
where we have made repeated use of \Eq{algebraic:identities:2}.
Thus, for each summand labeled by $q$ above, one is able to factor out the term $(x-y)$.
The general proof then follows by induction.

\clearpage
\newpage
\section{Analytic polynomials $\big(\,m_a(k),\; m_b(k)\,\big)$ for PNC in 
\Fig{fig:gmambpidiv2n2n3_BruteForceOnlyn2_mambofk}  and 
\Fig{fig:gmambpidiv3n2_analytic_solns_mambofk}}\label{tables:ma:mb:of:k}
In this Appendix we provide Tables II-IV, which list analytic results for zeros indicating complete destructive interference  for the case of a 50:50 BS ($\theta=\pi/2$: Table II for $n=2$, Table III for $n=3$), and for a non-50:50 BS ($\theta = \pi/3$: for $n=2$, Table IV).
\begin{table}[ht]
\centering
\begin{tabular}{|c|}
\hline
$\big(\,m_a(k),\; m_b(k)\,\big)$ \\
\hline
$(2 k^2 - k,\;              2 k^2 - 3 k + 1)$ \\
$(2 k^2 + k,\;             2 k^2 + 3 k + 1)$ \\
$(8 k^2 + k,\;            8 k^2 + 6 k + 1)$  \\
$(2 k^2 - 5 k + 3,\;     2 k^2 - 3 k + 1)$ \\
$(2 k^2 + 3 k + 1,\;     2 k^2 + 5 k + 3)$\\
$(8 k^2 + 6 k +1,\;    8 k^2 + 10 k + 3)$  \\
 $(2 k^2 + 5 k + 3,\;    2 k^2 + 7 k + 6)$ \\
$(2 k^2 + 7 k + 6,\;    2 k^2 + 9 k + 10)$ \\
\hline
\end{tabular}
\caption{Polynomials $(m_a(k), m_b(k))$ (with $m_b\ge m_a$) that identically solve $g(m_a,m_b,\theta=\pi/2|n=2)=0$;
see upper branch of \Fig{fig:gmambpidiv2n2n3_BruteForceOnlyn2_mambofk} (top).
Lower branch solutions are given by the pairs $(m_b(k), m_a(k))$.
}
\label{tbl:n2:theta:pidiv2}
\end{table}

\begin{table}[ht]
\centering
\begin{tabular}{|c|}
\hline
$\big(\,m_a(k),\; m_b(k)\,\big)$\\
\hline
  $(\quad\quad k \quad\quad,\;\qquad \quad\quad k \quad\quad)$\\
$(6 k^2 + 7 k + 2,\;    6 k^2 + 13 k + 7)$   \\
$(6 k^2 + 5 k + 1,\;    6 k^2 +11 k + 5)$ \\  
\hline
\end{tabular}
\caption{Polynomials $(m_a(k), m_b(k))$ (with $m_b\ge m_a$) that identically 
solve $g(m_a,m_b,\theta=\pi/2|n=3)=0$, (with  CNL $m_a = m_b = k$);
see the upper branch \Fig{fig:gmambpidiv2n2n3_BruteForceOnlyn2_mambofk} (bottom).
Lower branch solutions are given by the pairs $(m_b(k), m_a(k))$.
}
\label{tbl:n3:theta:pidiv2}
\end{table}

\begin{table}[ht]
\centering
\begin{tabular}{|c|}
\hline
$\big(\,m_a(k),\; m_b(k)\,\big)$\\
\hline
$(12 k^2 + k,\;            36 k^2 - 9 k )$   \\
$(12 k^2 + k,\;             36 k^2 + 15 k + 1)$   \\
$(12 k^2 + 7 k + 1,\;     36 k^2 + 9 k)$   \\
$(12 k^2 + 7 k + 1,\;     36 k^2 + 33 k + 7)$   \\
$(12 k^2 + 17 k + 6,\;     36 k^2 + 39 k + 10)$   \\
$(12 k^2 + 23 k + 11,\;     36 k^2 + 57 k + 22)$   \\
\hline
\end{tabular}
\caption{Polynomials $(m_a(k), m_b(k))$  that identically 
solve $g(m_a,m_b,\theta=\pi/3|n=2)=0$.
See plot legends of \Fig{fig:gmambpidiv3n2_analytic_solns_mambofk}.}
\label{tbl:n2:theta:pidiv3}
\end{table}

\section{On the HOM effect and analogue CNL for atomic systems}\label{app:HOM:CNL:Atoms}
The HOM effect depends upon the properties of the beam splitter coefficients for non-classical FS input states. 
Since these coefficients are essentially the components of the Wigner $SU(2)$ rotation matrices \cite{BAG_Parity:2021}, 
it is natural to ask if 
complete destructive interference of quantum amplitudes, and features analogous to the CNL, also exists for collections of atoms. The answer is yes if one makes use of the formal correspondence of the Schwinger representation for $su(2)$ in terms of a pair of bosonic modes \cite{Yurke:1986}:
\bsub
\bea{schwinger:su2}
J_x &=& \frac{1}{2}\, \left(a\,b^\dag + a^\dag\,b \right), \label{schwinger:su2:Jx} \\
J_y &=& \frac{i}{2}\, \left(a\,b^\dag - a^\dag\,b \right), \label{schwinger:su2:Jy} \\
J_z &=& \frac{1}{2}\, \left(a^\dag\,a + b^\dag\,b \right),  \label{schwinger:su2:Jz}
\eea
\esub
with commutators $\left[ J_i, J_j \right] = \epsilon_{i j k} J_k$, $i\in\{1,2,3\}\leftrightarrow\{x,y,z\}$.
In \Eq{schwinger:su2:Jx} (\Eq{schwinger:su2:Jy}), $J_x$ ($J_y$) has the form of a BS transformation of unitary form 
$U(\theta) =  e^{-i\,J_x\,\theta}\; \left( e^{-i\,J_y\,\theta}\right)$. Because of the use of the pseudo-angular momentum formalism involved in the Schwinger representation, a two-mode photon state 
$\ket{\Psi} = \sum_{n,m=0}^\infty c_{n,m}\,\ket{n,m}_{ab}$ can also be written in the angular momentum form
$\ket{\Psi} = \sum_{J=\{0,1/2,1,3/2\ldots\}}^{\infty} \sum_{M=-J}^J c_{J+M,J-M}\,\ket{J,M}_{ab}$ for angular momentum 
states $\ket{J,M}$ with angular momentum $J=(n+m)/2$ and spin projections $M=(n-m)/2\in\{-J, -J~+~1,\ldots,J\}$.

For $N_A$ two level atoms, the collective atomic spin operators
$J_i~=~\tfrac{1}{2}\, \sum_{k=1}^{N_A} \sigma^{(k)}_i$  define the 
\tit{Dicke} (pseudo angular momentum) states given by $\ket{J,M}$ with 
effective angular momentum $J=N_A/2$ and effective spin projections $M\in\{-J, -J~+~1,\ldots,J\}$.
The Dicke states $\ket{J,M}$, are defined in terms of the individual 
ground $\ket{g}$ and excited $\ket{e}$ atomic states as  $\ket{J,J}=\ket{\text{e}}^{\otimes N_A=2J}$ (all atoms excited) 
and  $\ket{J,-J}=\ket{\text{g}}^{\otimes N_A=2J}$ (all atoms in the ground state) 
with intermediate steps consisting of superpositions  of $J+M$ atoms in the excited state and $J-M$ atoms in the ground state. 

A general collective atomic (input) state (for a fixed number $N_A$ of atoms) is given by $\ket{\Psi_A} = \sum_{M=-J}^J\, c_M\, \ket{J,M}$.
The analogy of the 50:50 BS is the $\pi/2$ pulse of the Ramsey sequence acting on each angular momentum state
\bea{Ramsey}
\hspace{-0.15in}
\ket{J,M} \overset{U_{BS}}{\longrightarrow} \ket{J,M}^{(out)} &=& e^{-i\,J_x\,\theta}\ket{J,M} = 
\sum_{M'} \ket{J,M'}\,d^J_{M',M}(\theta), \qquad\;\;
\eea
for a general BS angle $\theta$.
Here, the Wigner rotation matrices $d^J_{M',M}(\theta)$ are related to the BS coefficients $f^{(n,m)}_p(\theta)$ via
\be{d:to:f}
d^{(n+m)/2}_{p-(n+m)/2, (n-m)/2}(\theta) \equiv f^{(n,m)}_p(\theta).
\ee
We can express each angular momentum state in terms of a pair of dual (Schwinger) $a,b$ number states (FS) via
\bsub
\bea{JM:nm}
\ket{J, M}_{JM} &\leftrightarrow&  \ket{n=J+M,m=J-M}_{ab} \\
 \ket{n,m}_{ab} &\leftrightarrow&\big|J=\frac{n+m}{2}, M=\frac{n-m}{2}\big\rangle_{J,M}.
\eea
\esub
Here $J+M$ is the number of atoms in their excited state and 
$J-M$ is the number of atoms in their ground state.
In the above we have put the subscript $JM$ on the atomic Dicke states to distinguish them from the $a,b$-mode FS.

With the above correspondences in hand, consider the general atomic input state for a fixed number of atoms $N_A=2 J$ given by
$\ket{\Psi_{in}}_A =$ $\sum_{n=0}^{2J}\,c_n \ket{J, M=n-J}_{JM}$ 
$= \sum_{n=0}^{2J}\,c_n \ket{n,m=2J-n}_{a,b}$ 
which is transformed by the BS transformation 
$U(\theta)=e^{-i\,J_x\,\theta}$ to 
$\sum_{n=0}^{2J}\,c_n\,\ket{n,m=2J~-~n}^{(out)}_{a,b}=$
$\sum_{n=0}^{2J}\,\sum_{p=0}^{2J}\,c_n\,f^{(n,2J-n)}_{p}(\theta)\ket{p, 2J-p}_{a,b}$.
If we now project this onto the dual FS $\ket{m_a,m_b}_{ab}$,
and consider only states with odd $n$, we formally have the output joint probability distribution 
\bsub
\bea{P:mamb:atoms}
\hspace{-0.65in}
P(m_a,m_b,\theta) &=& 
\left|
\sum_{n=0}^{2J} c_n \,f^{(n,2J-n)}_{p}(\theta)
\right|^2, \quad  \label{P:mamb:atoms:line1}\\
\hspace{-0.65in}
\overset{m_a=m_b=m'=J}{\underset{n\,odd}{\longrightarrow}}
P(m',m',\theta) &=& 
\left|
\sum_{n\,(odd)=0}^{2J} c_n \,f^{(n,2J-n)}_{J}(\theta)
\right|^2, \quad \label{P:mamb:atoms:line2} \\
\hspace{-0.65in}
&=&
\cos^2(\theta)\,
\left|
\sum_{n\,(odd)=0}^{2J} c_n \,\tilde{f}^{(n,2J-n)}_{J}(\theta)
\right|^2, \qquad\;\;  \label{P:mamb:atoms:line3}
\eea
\esub
where we have used $m_a+m_b=n+m=2J$ in \Eq{P:mamb:atoms:line1}
 and $m_a=m_b=m'=J$ in \Eq{P:mamb:atoms:line2}.
 Here, the summation $\sum_{n\,(odd)=0}^{2J}$
 indicates a sum over odd $n\in\{1,3,5,\ldots,n_{odd}^{(max)}\}$, 
 where $n_{odd}^{(max)}=2J-1\; (2J) $ if $J$ is an integer (half-integer).
 Finally, in \Eq{P:mamb:atoms:line2} we have used 
 $f^{(n,2J-n)}_{J}(\theta)\equiv \cos(\theta)\,\tilde{f}^{(n,2J-n)}_{J}(\theta)$ for \tit{each} $n$ odd, 
 (see \Eq{gmmtheta:n:odd} and \App{app:Proof}).
 Therefore, $P(m',m',\pi/2)=P(J,J,\pi/2)~=~0$ for a $50:50$ BS ($\theta=\pi/2$) \tit{regardless} of the coefficients $c_n$, as long as $n$ is odd. 
 
 Now the projection onto the dual FS $\ket{m',m'}_{ab} = \ket{J~=~m',M~=~0}_{JM}$ is a projection onto
 the atomic state with equal number of atoms in the 
 ground and excited states for a collection of $N_A = 2J = 2m'$ atoms.
The state $\ket{J,M=0}$ can only occur, of course, for integer $J\in\mathbb{Z}_{\ge 0}$. 
 \Eq{P:mamb:atoms:line3} shows that the quantum amplitude for this state will suffer complete destructive interference for a 50:50 BS ($\cos(\pi/2)=0$) and hence will not be present in the output state (the HOM effect). 
For half-integer values of $J$ there is simply never the possibility of an output state from the BS of equal number of atoms in both their excited and ground states (see the similar discussion after \Eq{gmmtheta:n:even} for photons), and hence no quantum interference effect is taking place.
 
 With respect to the concept of CNL (as in the photonic case) we will also find that
 for $\tilde{m}\ne m'$, $P(\tilde{m},\tilde{m},\pi/2)~=~0$, but this will now entail having to make a measurement on a new number of atoms given by $\tilde{N}_A = 2\tilde{J}=2\tilde{m} \ne N_A$. This means that the atomic ``CNL" is 
 \tit{spread out} over a collection of measurements of experiments involving a different number of atoms for each diagonal element of the output probability distribution $P(m',m',\pi/2)=P(J,J,\pi/2)$,
 i.e. it is not a \tit{contiguous} set of zeros representing complete destructive interference for a \tit{single} collection of atoms $N_A$. However, for a given fixed number of atoms $N_A=2J$, with input states of the form $\sum_{n\,(odd)=0}^{2J} c_n \ket{J,M=n-J}_{JM}$ we have $P(J,J,\pi/2)=0$, which is the extended HOM effect.
 Once again, (see the discussion after \Eq{gmmtheta:n:even}) half the points in the atomic ``CNL" (spread out over different number of atoms $N_A = 2\,J$) are trivially zero, since the state $\ket{J,M=0}$ is simply not capable of being present in the output (for $J$ half integer), 
 while the other half of points (for $J$ integer) correspond to diagonal output states that have coefficients proportional to 
 $\cos(\theta)\overset{\theta=\pi/2}{\longrightarrow} 0$ for a 50:50 BS, which is the extended HOM effect of complete destructive interference of the output amplitude for the state $\ket{J,M=0}$ (i.e. equal number of atoms in the excited and ground state). This is the atomic analogue of the CNL in the photonic case where the extended HOM effect occurs on output states 
 $\ket{m'=(n+m)/2,m'}_{ab}$ with equal number of photons in each mode.

\end{document}